\documentclass[letterpaper,showpacs,onecolumn,eqsecnum,superscriptaddress,
floatfix]{revtex4}

\usepackage{amssymb}
\usepackage{amsmath}
\usepackage{latexsym}
\usepackage{mathtools}
 
\usepackage{tensor} 
\usepackage{framed}
\makeatletter
\newcommand{\mathleft}{\@fleqntrue\@mathmargin0pt}
\newcommand{\mathcenter}{\@fleqnfalse}
\makeatother
\usepackage{epsfig}
\usepackage{graphicx}
\usepackage{subcaption}
\usepackage{color}
\usepackage[linktocpage]{hyperref}

\begin{document}

\title{Geodesic structure of a rotating regular black hole}

\author{Brandon Bautista-Olvera} 
\email[]{bbautista@icf.unam.mx}
\affiliation{Instituto de Ciencias F\'isicas, Universidad Nacional Aut\'onoma 
de M\'exico, Apartado. Postal 48-3, 62251, Cuernavaca, Morelos, M\'exico}

\author{Juan Carlos Degollado} 
\email[]{jcdegollado@icf.unam.mx}
\affiliation{Instituto de Ciencias F\'isicas, Universidad Nacional Aut\'onoma 
de M\'exico, Apartado. Postal 48-3, 62251, Cuernavaca, Morelos, M\'exico}

\author{Gabriel German} 
\email[]{gabriel.german@icf.unam.mx}
\affiliation{Instituto de Ciencias F\'isicas, Universidad Nacional Aut\'onoma 
de M\'exico, Apartado. Postal 48-3, 62251, Cuernavaca, Morelos, M\'exico}

\date{\today}

\bigskip

\begin{abstract}

We examine the dynamics of  particles around a rotating regular black hole.
In particular we focus on the effects of the characteristic length parameter of the spinning black hole 
on the motion of the particles by solving the equation of orbital motion.
We have found that there is a fourth constant of motion that determines the dynamics of orbits out the equatorial plane similar as in the Kerr black hole.
Through detailed analyses of the corresponding effective potentials for massive particles the possible orbits are numerically simulated. A comparison with the trajectories in a Kerr spacetime shows that the differences appear when the black holes rotate slowly for large values of the characteristic length parameter.

\end{abstract}



\maketitle

\section{Introduction}
\label{sec:introduction}
Black holes are perhaps the most fascinating and mysterious objects in the 
Universe. Although  classical black holes are well described by General 
Relativity they present a singularity where the theory is unable to give a 
complete description. Singularities thus, are regarded as 
indicating the breakdown of the theory. It has been argued that quantum 
modifications of general relativity may correct the singular behavior of black 
hole solutions and a complete 
description of quantum gravity will circumvent these problems, however, such  
formulation is still missing. 

Although singularities are hidden by event horizons thus being 
harmless for the description of physics outside black holes, studying regular 
black holes is a problem of interest.
Several attempts have been made to eliminate the singularity on black holes by 
replacing it at 
the Planck scale curvature by the de Sitter geometry. For instance, effective 
theories introduce a minimal cut-off length, close to the Planck length, 
yielding a singularity-free space time.
There are also some classical approaches that cure the singularity of the 
spacetime by coupling gravity to an external form of matter, sometimes modeled 
by some form of nonlinear electrodynamics. 
The first regular black hole 
solution 
was reported by Bardeen in~\cite{Bardeen:1972fi} and since then, many metrics 
that
are spherically symmetric, static, asymptotically flat and have regular centers 
have been found~\cite{BarrabesFrolov1996, Marsetal1996, CaboAyon-Beato1999, bron2,
 Dymni92, Dymnikova:2004zc, Balart:2014cga}. Ayon-Beato and Garc\'ia~\cite{AyonBeatoGarcia98} and later Burinskii and Hilderbrandt~\cite{Burinskii02} showed that the Bardeen solution
 represents a black hole with a magnetic monopole in a nonlinear electrodynamics.

Among the most noticeable models of regular black holes was the spherically symmetric spacetime proposed 
by Hayward in 2006~\cite{Hayward:2005gi}.
Fan and Wang showed in~\cite{Fan:2016rih,Fan:2016hvf}
that this spacetime results from Einstein's equations 
in the presence of a magnetic charge
in a nonlinear electromagnetic field as a source. 
Later, rotating versions of the Hayward metric were 
introduced by Bambi and Modesto~\cite{Bambi:2013ufa}.
The metrics derived by Bambi and Modesto were constructed using the Newman-Janis algorithm commonly used to produce axisymmetric spacetimes from a  spherically symmetric solution,  see~\cite{Drake:1998gf} for a detailed derivation.
The resulting metric still represents a regular black hole, nevertheless it was shown in~\cite{Toshmatov:2017zpr, Rodrigues:2017tfm, Toshmatov:2017anu} 
that the matter content needed to produce the geometry is 
only an approximation to its corresponding Hayward nonrotating magnetic monopole source.

The interior of a black hole is a hidden region by definition.  The surface 
that connects the interior to the exterior is the horizon.
Thus, a way to peer into the interior is to study the near horizon region.
While the study of the physical properties of black holes represents a wide 
field, the most basic study to be performed is a study of the geodesics in 
these spacetimes. The motion of test particles around black holes has been 
studied extensively over the years. For most of the common black holes, e.g. 
Schwarzschild, Reissner-Nordstr\"om, or Kerr, analytic solutions to the 
equations of motion can be given in terms of elliptic functions 
\cite{Hackmann:2008zza, Hackmann:2010zz}. For a non rotating Hayward
black hole a geodesic study in the equatorial plane was made by Abbas and Sabiullah in~\cite{Abbas:2014oua}.
The geodesic equations for more general black holes become complicated and solutions to the 
orbits in a closed form are unavailable. Still, the symmetries of the spacetime 
may be used to simplify the geodesic equations and provide relevant information 
about the motion of particles around the black hole. The purpose of the present 
work is to analyze the motion of massive test particles in the vicinity of a 
rotating regular black hole. We use effective potential methods to characterize 
the motion and present numerical solution for the equations of motion out of the 
equatorial plane. These results are of particular interest if one considers for 
instance accretion processes. Usually, studies of accretion focus only on 
particles moving in the equatorial plane, because this is the easiest case, but 
particles initially moving around some particular orbit may be perturbed by a 
kick along the $\theta$ direction and give rise to a much richer orbits. 
Furthermore, studies of geodesics out of the equatorial plane are important from the gravitational wave point of view. In a Kerr spacetime for instance, the evolution of the trajectories
is entirely determined by the radiated energy and angular momentum. 

One can expect that astrophysical black holes are rotating. A progenitor 
massive star has non vanishing angular momentum. Even if some part of angular 
momentum is lost during the the black hole formation process, the resulting 
black hole would be rotating.
Up to now the Kerr solution of the Einstein's equation has played a major role 
in the description of astrophysical black holes \cite{Valtonen:2008tx}.
The recent observations with the Event Horizon Telescope  have provided us with an image of the shadow of the super-massive black hole M87 which is consistent with the shadow of a Kerr black hole.
The study of alternatives to the Kerr solution as the one presented 
here  and the study of geodesics around regular rotating black holes, 
is a useful test bed for exploring minimal departures from classical black hole 
geometries, additionally such studies may contribute to a better understanding of the data coming in the future~\cite{Akiyama:2019cqa}.

The paper is organized as follows:
In section~\ref{sec:spacetime} we present some basic properties of the 
rotating Hayward metric proposed by Bambi and Modesto~\cite{Bambi:2013ufa}, and 
present a characterization of the horizons and the ergospheres.
In section~\ref{sec:geodesics} we derive the equations of motion using the Hamilton Jacobi approach paying particular attention on the conserved quantities and in the separability of the equations. We also present some 
properties of radial and polar motions using the effective potentials and highlight some features of the 
radial and angular motion. In section~\ref{sec:trajectories} we show the numerical solutions of the equations of motion and discuss the trajectories of the particles in the configuration space.
Finally, in section~\ref{sec:Conclusions} we discuss the results and present some concluding remarks.
Through the paper we use geometrical units with $G = c = 1$.

\section{Spacetime properties}
\label{sec:spacetime}
The spherically symmetric metric proposed by Hayward~\cite{Hayward:2005gi} is 
given by
\begin{equation} \label{eq:metric_hayward}
 ds^2 = -f(r)dt^2 + \frac{1}{f(r)}dr^2 + r^2 (d\theta^2+\sin^2\theta 
d\phi^2) \ , 
\end{equation}
with 
\begin{equation}
f(r):=1-\frac{2m r^2}{r^3+2m \ell^2} \ ,
\end{equation}
where $m$ is the Arnowitt-Desser-Misner mass and the parameter $\ell$, is of 
the order of the Planck length,
for $\ell =  0$ the metric~(\ref{eq:metric_hayward}) reduces to Schwarzschild. 
The Hayward spacetime is asymptotically flat and  in the limit $r\rightarrow0$ ~(\ref{eq:metric_hayward}) behaves as a de Sitter metric with a cosmological constant $\Lambda=\frac{3}{\ell^2}$~\cite{Hayward:2005gi}. Fan and Wang have shown that the Hayward metric can 
be 
obtained as a solution of Einstein's equations coupled  with a nonlinear 
electrodynamics with a magnetic monopole as a source~\cite{Fan:2016hvf}.  
Depending on the relative values of $m$ and $\ell$ the 
metric~(\ref{eq:metric_hayward}) can have one, two or zero horizons.

By using the Newman-Janis algorithm Bambi and Modesto~\cite{Bambi:2013ufa} 
obtained a family of possible generalizations of Hayward metric to include 
rotation. In this work we will focus on a metric that has the same form as the 
Kerr metric with a varying mass.
\begin{equation}
 ds^2 =-\left( 1-\frac{2 r M(r)}{\Sigma}  \right) dt^2 - \left(\frac{4M(r)ar 
\sin^2\theta}{\Sigma} \right)dtd\phi + \frac{\Sigma}{\Delta}dr^2 +\Sigma 
d\theta^2 +
\left(r^2+a^2 +\frac{2M(r)a^2r \sin^2\theta}{\Sigma}\right)\sin^2\theta 
d\phi^2 \ ,
\label{eq:metric_rothayward}
\end{equation}
where the functions $\Delta$, $\Sigma$ and $M$  are defined as:
\begin{equation}\label{eq:coefmetric_rothayward}
 \Delta := r^2 -2M(r)r +a^2,  \qquad {\rm} \qquad \Sigma:= r^2+a^2\cos^2\theta, 
\qquad M(r):=\frac{m r^2}{r^3+g^3} \ .
\end{equation}
The length parameter $g$, is of the order of the Planck length and is a measure of the deviation of the Kerr spacetime.
The components of the metric inverse, which will become useful later, are

\begin{eqnarray}\label{eq:inmetric_rothayward}
&g^{tt} = - \frac{1}{\Delta \Sigma}\left[ (r^2+a^2)^2-a^2\Delta \sin^2\theta 
\right] \ , \qquad g^{t\phi} = -\frac{2M(r)ar}{\Delta \Sigma}\ , \qquad 
g^{rr}=\frac{\Delta}{\Sigma} \ ,&\\ \nonumber
&g^{\theta\theta} = \frac{1}{\Sigma}\ , \qquad g^{\phi\phi} = 
\frac{\Delta -a^2\sin^2\theta}{\Delta\Sigma \sin^{2}\theta}\ . &
\end{eqnarray}

The spin parameter $a$ is related with the total 
angular momentum $J$, by $a=J/m$.
For $g=0$ equations~(\ref{eq:metric_rothayward}) 
and~(\ref{eq:inmetric_rothayward}) reduce to the Kerr solution in 
Boyer-Lindquist coordinates.
As described in~\cite{Bambi:2013ufa} the spacetime 
metric~(\ref{eq:metric_rothayward}) is regular at $r=0$ for $g 
\neq 0$. A rigorous analysis about the regularity of this spacetime can be found in Torres and Fayos~\cite{Torres:2016pgk}. In particular, it 
has been argued that an extension of the spacetime with values $r< 0$ does 
present a singularity~\cite{Lamy:2018zvj}. To avoid such singularity 
we focus in the region covered by $r\geq 0$.

\subsubsection{Horizons}

\begin{figure}[ht!]
\begin{center}
\includegraphics[scale=0.32]{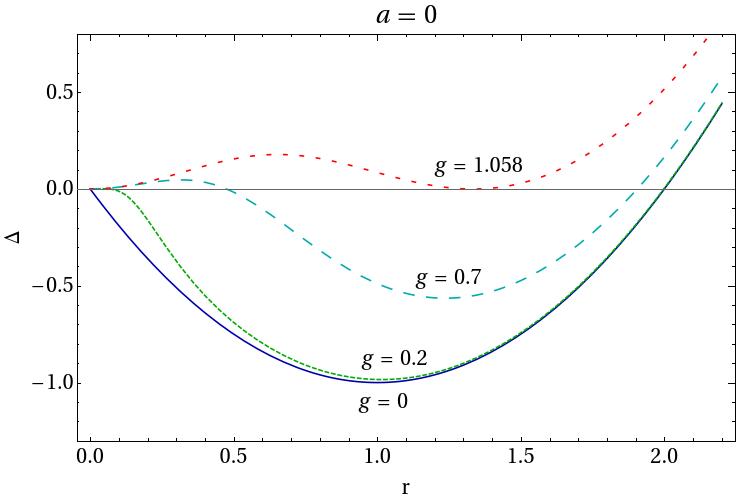}
\includegraphics[scale=0.32]{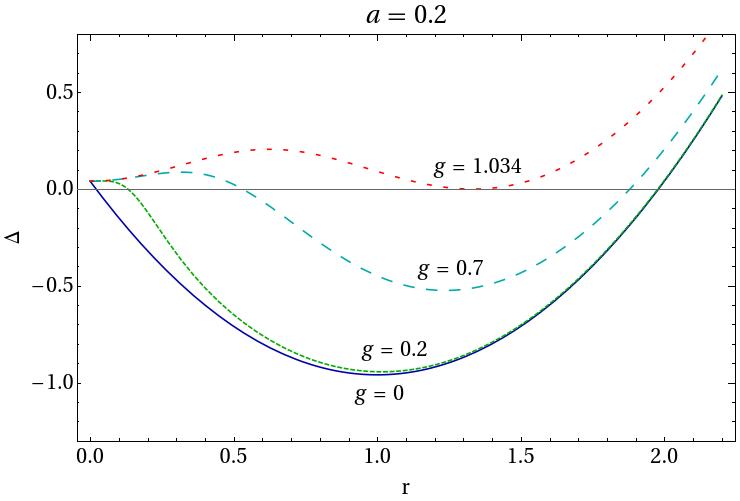}
\includegraphics[scale=0.32]{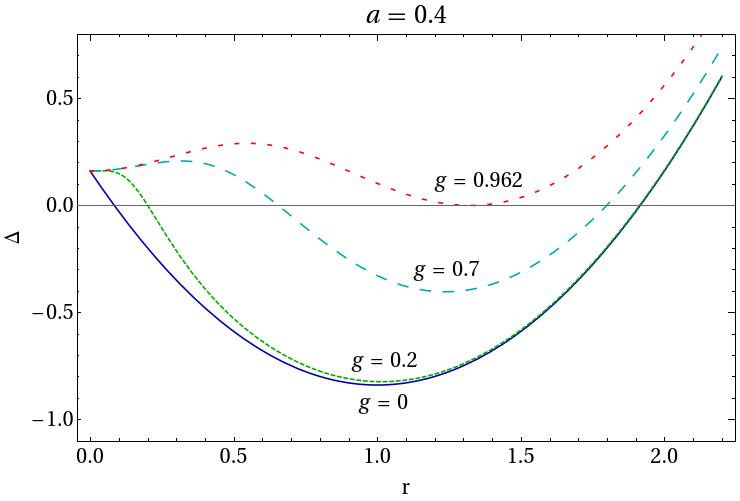}
\includegraphics[scale=0.32]{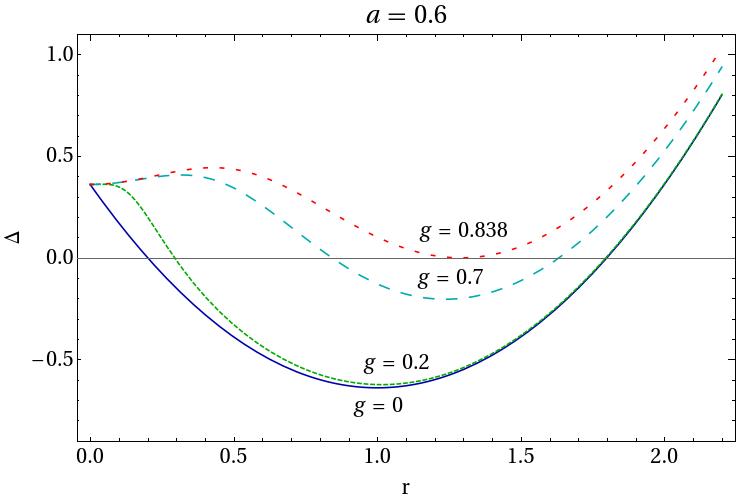}
\includegraphics[scale=0.32]{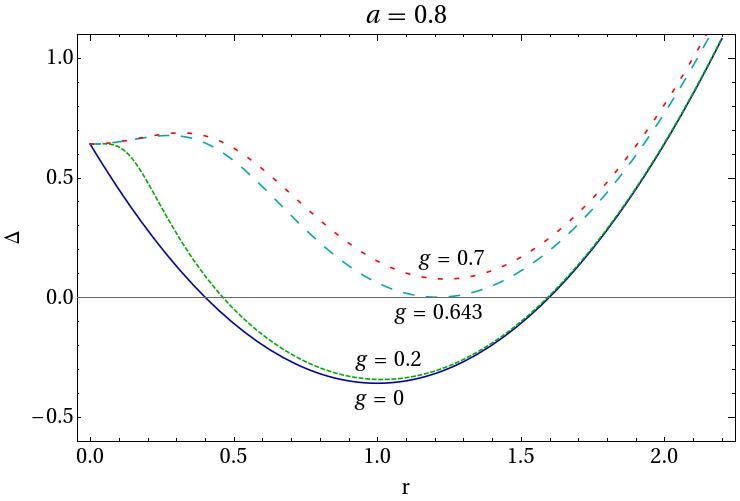}
\includegraphics[scale=0.32]{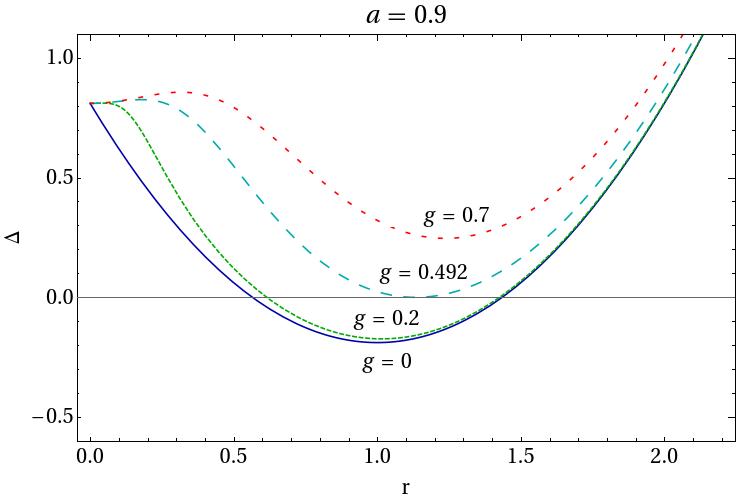}
\caption{Possible spacetime structures for rotating regular black holes 
depend on the values of $g$ and $a$.
The zeroes of $\Delta$ determines the presence of two, one or none horizons.
}
\label{fig:horizon}
\end{center}
\end{figure}

\begin{figure}[ht!]
\begin{center}
\includegraphics[scale=0.4]{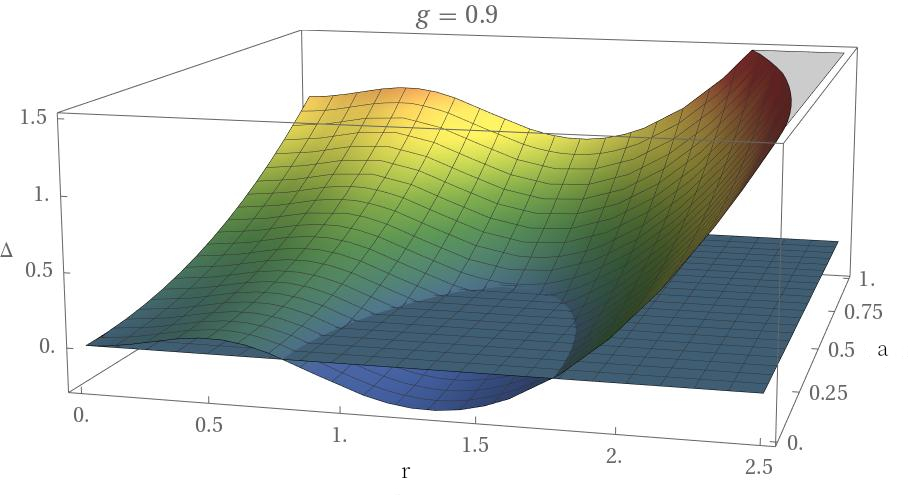}
\caption{Surface $\Delta$ as a function of $r$ and $a$ for $g=0.9$. The intersection of the surface with the plane $z=0$ for a 
given value of $a$ determines the position of the horizons.}
\label{fig:horizon3d}
\end{center}
\end{figure}

The horizons are defined by the relation 
$ g_{rr} \rightarrow \infty$ which is equivalent to set
\begin{equation}\label{eq:horizons}
 \Delta = r^2-2M(r)r+a^2=0 \ .
\end{equation}
This equation allows for real solutions for the radius depending on the values of the 
parameters $a$ and $g$.
The outermost radius determines the event horizon location. In the limit 
$g\rightarrow 0$ there are two horizons for 
values 
$0<a<m$. In the extreme case, $a=m$ the two horizons coincide.  
For the nonrotating case, the equation defining the location of the horizons 
becomes
\begin{equation}
 1-\frac{2mr^2}{r^3+2m\ell^2} = 0 \ ,
\end{equation}
which implies a critical mass $m_c=3\sqrt{3}\ell/4$ and a critical radius 
$r_c=\sqrt{3}\ell$, such that the horizons are degenerate at $r=r_c$ when 
$m=m_c$.
When $m<m_c$ there are two horizons whereas there are no horizons when $m>m_c$ 
\cite{Hayward:2005gi}. When $a\neq 0$ Eq.~(\ref{eq:horizons}) becomes a 
polynomial of fifth order in $r$ and it becomes necessary to find the roots numerically.  
Fig.~\ref{fig:horizon} shows the behavior of  
$\Delta$ as a function of $r$ for several values of the spin parameter $a$. The 
intersection with the horizontal axis determine the position of the horizons, and depending on the relative values of $a$ 
and $g$ the spacetime posses, two, one or none horizons. 
The left plot in the first row of  Fig.~\ref{fig:horizon} displays the behaviour of $\Delta$ for the non rotating case $a=0$. For values $g<1.058$ the spacetime has two horizons, for the threshold value $g=1.058$ the spacetime has one horizon and no horizons are present for $g>1.058$. For positive values of the spin $a$ the threshold value of $g$ changes as shown in the rest of the plots in Fig.~\ref{fig:horizon}. For instance, in the right plot of the third row for a spacetime with $a=0.9$, the threshold value is $g=0.492$. It can be seen that, as the value of $a$ approaches to 1, the threshold value of $g$ decreases.
Further information can be obtained by keeping the value of $g$ fixed.
Fig.~\ref{fig:horizon3d} shows a surface plot of $\Delta$ as a function of 
$a$ and $r$ for $g=0.9$. From the figure one can see that,
as the value of $a$ increases from zero to a critical value $a_c<1$, two horizons are present. For values of $a$ between $a_c$ and one there are no horizons. 
Remarkably and unlike the Kerr case, the degeneration of the horizon (the extreme black hole) happens for values of $a$ less than 
one.

\subsubsection{Ergosphere}
\begin{figure}[h!]
\begin{center}
\includegraphics[scale=0.21]{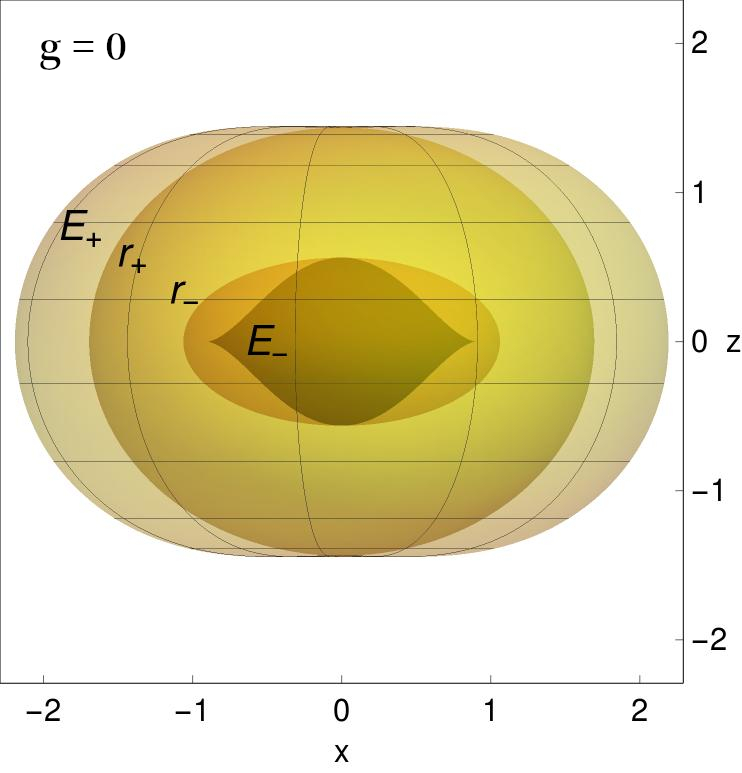}
\includegraphics[scale=0.21]{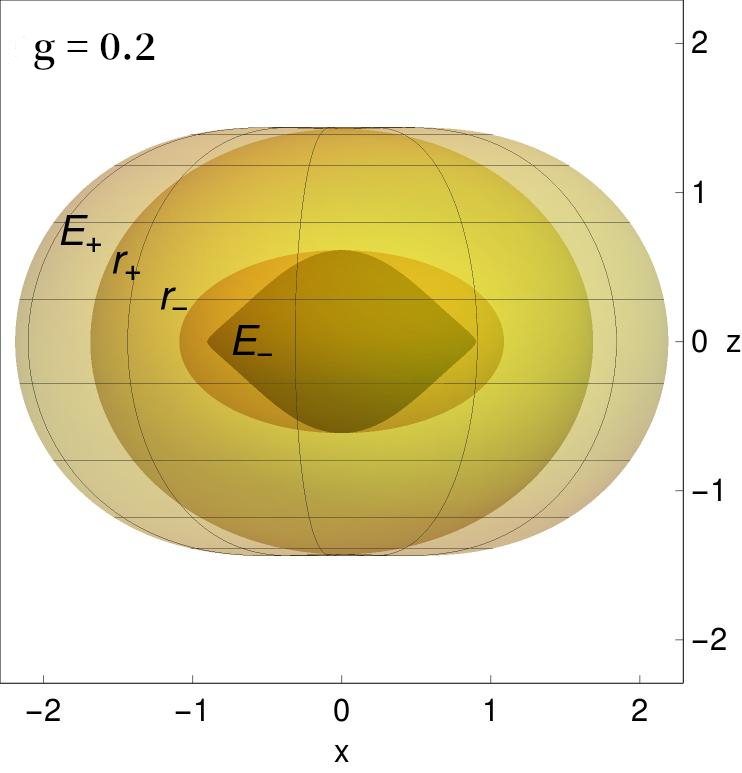}
\includegraphics[scale=0.21]{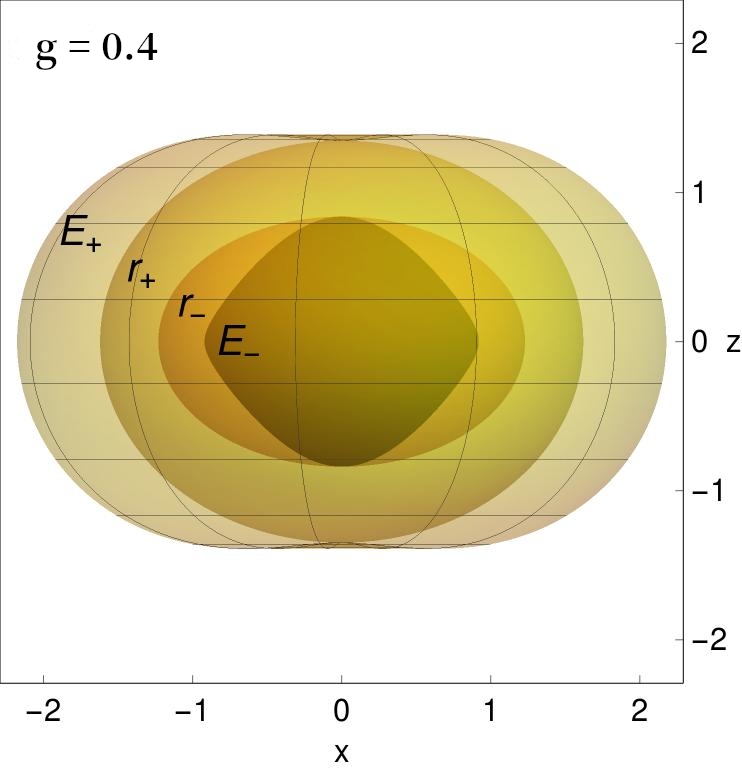}
\includegraphics[scale=0.21]{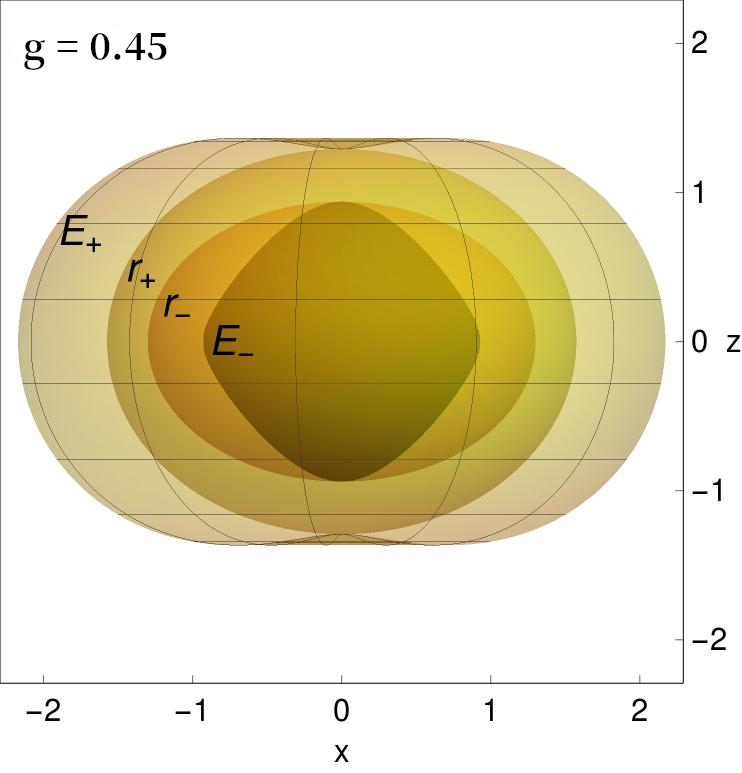}
\includegraphics[scale=0.21]{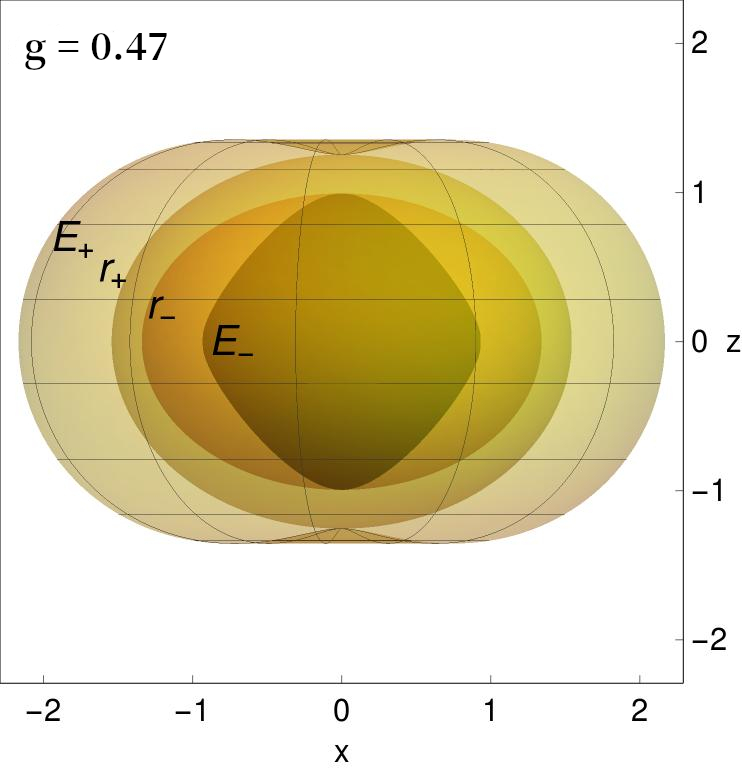}
\includegraphics[scale=0.21]{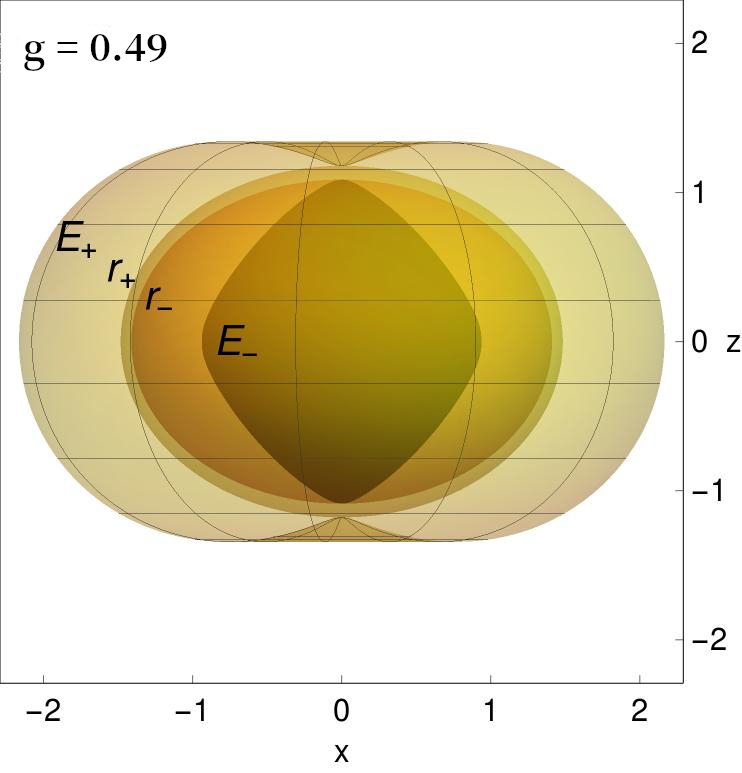}
\caption{The ergoregion is delimited by the limit surface defined by 
equation~(\ref{eq:egosphere}). The Kerr black hole corresponds to $g=0$.
The rotation parameter used in the plots is $a = 0.9$. The effect of the 
parameter $g$ is more dramatic in the inner horizon $r_{-}$ and in the 
inner ergosphere $E_{-}$.
}
\label{fig:ergosphere}
\end{center}
\end{figure}

There is another important surface of rotating black holes. The static 
limit. When a particle crosses the static limit the nature of the particle 
changes; a timelike geodesic becomes spacelike and a spacelike geodesic becomes timelike.
The static limit surface is defined by the equation
\begin{equation}\label{eq:egosphere}
g_{tt} = r^2+a^2\cos^2\theta-2M(r)r=0 \ . 
\end{equation}
The ergosphere is a region located between the event horizon and the static 
limit surface. In this region particles can extract energy from the black hole 
via the Penrose process~\cite{Penrose:1964wq, Pourhassan:2015lfa}. A throughout 
study of the 
ergoregion of a rotating Hayward BH 
was performed in~\cite{Amir:2016nti}. The authors conclude that the ergoregion is 
enlarged compared to Kerr as the value of the parameter $g$ increases. 
Fig.~\ref{fig:ergosphere} illustrates the behavior of the shape and extension 
of the ergosphere of 
the Hayward rotating black hole
for various values of the rotation parameter $a$ and the parameter $g$.
In order to plot these surfaces, we have used the spheroidal-like coordinates
\begin{eqnarray}
 x&=&\sqrt{r^2+a^2}\cos\phi\sin\theta \ ,\\
 y&=&\sqrt{r^2+a^2}\sin\phi\sin\theta \ ,\\
 z&=&r\cos\theta \ .
\end{eqnarray}
In Fig.~\ref{fig:ergosphere} we show a sequence of ergoregions and horizons 
of the rotating Hayward black hole varying the parameter $g$.
The labels $r_{+}$ 
and $r_{-}$ at the surfaces in Fig.~\ref{fig:ergosphere} denote the outer and 
inner 
horizons, while $E_{+}$ and $E_{-}$ represent the outer and 
inner ergospheres.
The plots correspond to values of $a = 0.9$ and $g$ running from 
$g=0$ to $g=0.49$. Note that, as the value of $g$ increases the two 
horizons merge into one.

In order to give a quantitative measure of the effect of $g$ on the ergospheres 
and horizons
we compare the ratio between the length of the geodesic between the north and 
south pole 
and the equatorial 
circle 
$\varepsilon(g;R) := \frac{l_p}{l_e}$,
where
\begin{eqnarray}
l_p &=& \int_{0}^{\pi} (R^2 + a^2 \cos\theta)^{1/2} d\theta \ ,
\end{eqnarray}
and 
\begin{eqnarray}
l_e &=& \int_{0}^{2\pi}\left(R^2 + a^2 + \frac{2a^2 
M(R)}{R}\right)^{1/2} d\phi \ .
\end{eqnarray}
The surface $R$ will take the values 
$r_{+}$, $r_{-}$, $E_{+}$ $E_{-}$ respectively.

\begin{figure}[h!]
\begin{center}
\includegraphics[width=0.5\textwidth]{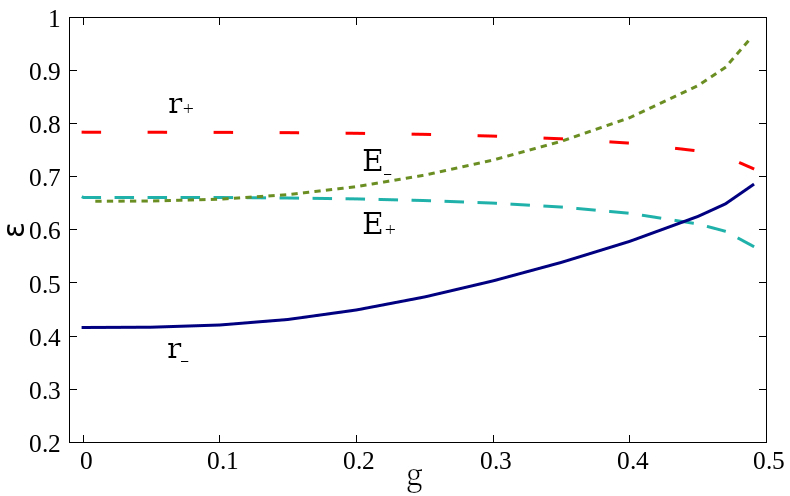}
\caption{The ratio between the length of the pole to pole geodesic and the 
equatorial circle $\varepsilon(g;R)$ for a rotation parameter 
$a=0.9$ for the outer(inner) horizons and outer(inner) ergospheres.}
\label{fig:eccentricityg}
\end{center}
\end{figure}

Fig.~\ref{fig:eccentricityg} shows the ratio $\varepsilon$ as a 
function of $g$ for the horizons $r_{\pm}$ and ergospheres $E_{\pm}$.
One can see that $\varepsilon(g;r_{+})$ and $\varepsilon(g;E_{+})$ decreases 
as $g$ grows, while  $\varepsilon(g;r_{-})$ and $\varepsilon(g;E_{-})$ are 
increasing functions. The Kerr case in this plot would correspond to 
constant horizontal lines at the value $\varepsilon(0;R)$.
What we can infer from the ratio $\varepsilon(g;R)$ is that the oblate surfaces 
$r_{+}$ and $E_{+}$ get extended at the equator while the surfaces $r_{-}$ and 
$E_{-}$ become narrower and extended at the poles in accordance with the 
behavior shown in Fig.~\ref{fig:ergosphere}.

%
\section{Geodesic motion of timelike particles}
\label{sec:geodesics}

In order to give a general outlook of the geodesic motion of massive particles in the rotating Hayward spacetime, we use  three different but otherwise  equivalent formalisms each one suitable to show a specific aspect of the motion. First, we appeal to the Hamilton-Jacobi formalism to show that a fourth integral of motion associated to the Carter constant exist and  the problem is completely integrable. Then, once we have proven the motion is separable into radial and polar parts we
describe the motion by means of 
effective potentials. Finally in order to integrate numerically the equations, we use the hamiltonian formalism.

\subsubsection{Hamilton Jacobi approach}

In the stationary and axially symmetric Kerr spacetime the geodesic equations 
are completely integrable.
There are two 
obvious conserved quantities given by the symmetries of the spacetime; the 
energy and the azimuthal angular momentum. Furthermore, the condition 
$p^{\mu}p_{\mu}= - \mu^2$ gives another integral of motion and a fourth integral 
 was discovered by 
Carter~\cite{Carter:1968rr}. This fourth integral can be obtained by separating 
variables in the Hamilton Jacobi equation 
of motion. 

Let us consider the geodesic motion of a massive test particle 
in the background of the 
metric~(\ref{eq:metric_rothayward}). The Hamilton-Jacobi equation for this 
particle has the form:
\begin{equation}\label{eq:ham-jacobi}
\frac{1}{2} g^{\alpha\beta}\frac{\partial S}{\partial x^\alpha}\frac{\partial 
S}{\partial 
x^\beta}+\frac{\partial S}{\partial \tau}=0 \ ,
\end{equation}
where 
$\tau$ is the proper time. 
%
Because the metric does not depend on $t$ nor $\phi$, we can introduce 
two integrals of motion along the trajectories, namely the energy
$E$ and the axial angular momentum $L_z$ as: 
\begin{eqnarray}
 p_{t} &=&\frac{\partial S}{\partial t} = -E \ , \\ 
 p_{\phi} &=&\frac{\partial S}{\partial \phi} = L_z \  .
\end{eqnarray}
As in the Kerr black hole, 
the Hamilton Jacobi function $S$ for a geodesic in the Hayward rotating black 
hole can be written in the following separated form 
\begin{equation}\label{eq:ansatz_separ}
 S=\frac{1}{2} \tau -Et+L_z\phi + S_{r}(r) + S_{\theta}(\theta)\ ,
\end{equation}
where we have set the mass of the particle $\mu=1$.
Substituting the components of the inverse
metric~(\ref{eq:inmetric_rothayward}), the ansatz (\ref{eq:ansatz_separ})  and the corresponding momenta into Eq. ~(\ref{eq:ham-jacobi}) we obtain:
\begin{eqnarray}\label{eq:separability}
 -\left(  \frac{(r^2+a^2)^2 }{\Delta}-a^2\sin^2\theta  \right)E^2 + 
\frac{4M(r) a r}{\Delta} EL_z + \left( 
\frac{1}{\sin^2\theta}-\frac{a^2}{\Delta}  \right)L_{z}^2 + \Delta 
\left(\frac{ d S_r}{d r}\right)^2 +  
\left(\frac{ d S_{\theta}}{d \theta}\right)^2+ \Sigma = 0 ,
 \end{eqnarray}

Simplifying the equation~(\ref{eq:separability}), introducing a separation 
constant $k$ representing an additional constant of motion, and the 
Carter's constant through the relation $Q = k - (aE-L_z)^2$, we arrive at the 
following differential equations for $S_r(r)$ and $S_\theta(\theta)$:
\begin{eqnarray}
 \Delta \left(  \frac{d S_r}{dr}  \right)^2 = \frac{R(r)}{\Delta} \ , \qquad 
{\rm and}\qquad
 \left(  \frac{d S_{\theta}}{d\theta}  \right)^2 = \Theta(\theta)\ , 
\end{eqnarray}
where
\begin{eqnarray}
 R(r) &=& \left[ (r^2+a^2)E - aL_z \right]^2 -\Delta \left[ (aE - L_z)^2+r^2+Q
\right] \ , \\
\Theta(\theta)&=& Q -\left[ \frac{L_z^2}{\sin^2\theta} +a^2(1 - E^2)  
\right]\cos^2\theta \ .
\end{eqnarray}
The resulting expressions for the radial and angular momentum are
\begin{equation}\label{eq:pr_and pth}
 p_r = \frac{\partial S}{\partial r} = \frac{\sqrt{R}}{\Delta}\ , \qquad 
 p_{\theta}=  \frac{\partial S}{\partial \theta}= \sqrt{\Theta}\ .
 \end{equation}
Formally, the solution of the Hamilton Jacobi equation is the principal function 
$S$ of the form:
\begin{equation}
 S= \frac{1}{2}\tau-Et + L_z \phi + 
\int_{0}^{r}\frac{\sqrt{R(r')}}{\Delta}dr' + \int^{\theta}_0 \sqrt{\Theta 
(\theta')}d\theta' \ .
\end{equation}

While this solves the problem theoretically, it is difficult to understand the 
orbit structure analytically and numerical solutions become necessary.

In order to get a 
relation between coordinates and momenta
we use the lagrangian:
\begin{equation}
 \mathcal{L}= \frac{1}{2}\,g_{\alpha\beta}\,\dot x^{\alpha}\dot x^{\beta}\  ,
\end{equation}
and thus
\begin{equation}
 p_{\alpha}=\frac{\partial  \mathcal{L}}{\partial \dot x^{\alpha}} = 
g_{\alpha\beta}\dot x^{\beta} \ ,
\end{equation}
where dot 
denote derivation with respect to the proper time $\tau$. 

Explicitly, by the spacetime under study
\begin{eqnarray}\label{eq:ps_lagrange}
p_{t}&=&\frac{\partial\mathcal{L}}{\partial\dot t} = 
-\left(1-\frac{2M(r)r}{\Sigma}\right)\dot t-\frac{2M(r)ar\sin^2\theta}{\Sigma} \dot 
\phi \ , \\ \nonumber
p_{r} &=& \frac{\partial\mathcal{L}}{\partial\dot r} =  
\frac{\Sigma}{\Delta}\dot r \ , \\ \nonumber
p_{\theta} &=& \frac{\partial\mathcal{L}}{\partial\dot \theta} =  
{\Sigma}\dot \theta \ ,\\ \nonumber
p_{\phi} &=& \frac{\partial\mathcal{L}}{\partial\dot \phi} = 
-\frac{2M(r)ar\sin^2\theta}{\Sigma} \dot t + \left( 
(r^2+a^2)+\frac{2M(r)a^2r\sin^2\theta}{\Sigma}  \right) \sin^2\theta \dot\phi \ .
\end{eqnarray}
One can thus, by using Eqs.~(\ref{eq:pr_and pth}) and (\ref{eq:ps_lagrange}), write the equations of the geodesic motion in the first 
order form in the lagrangian formalism as
\begin{eqnarray}
\Sigma \frac{dr}{d\tau} &=& \sqrt{R(r)} \ , \label{eq:forder_r} \\
\Sigma \frac{d\theta}{d\tau} &=& \sqrt{\Theta(\theta)} \ ,\label{eq:forder_th} 
\\
\Sigma \frac{d\phi}{d\tau} &=& -\left(aE-\frac{L_z}{\sin^2\theta} 
\right)+\frac{a}{\Delta}\left(E(r^2+a^2)-aL_z \right) \ , \label{eq:forder_ph} 
\\
\Sigma \frac{dt}{d\tau} &=& -a\left(aE{\sin^2\theta}-{L_z} 
\right) +\frac{r^2+a^2}{\Delta}\left[E(r^2+a^2)-aL_z \right]\ . 
\label{eq:forder_t}
\end{eqnarray}
%
However, equations (\ref{eq:forder_r}) and  (\ref{eq:forder_th})
present terms with square roots and  it is known that these terms causes difficulties in numerical solutions due to the change of 
signs in the turning points, see for instance~\cite{Fuerst:2004ii}. For this reason becomes preferable to reformulate the equations of motion using other equivalent approaches.

Before going deeper into the solutions of the equations of motion one can split the analysis considering the motion in $r$ and $\theta$
separately and get some insight about the general motion. 

In the following we show that equations (\ref{eq:forder_r}), (\ref{eq:forder_th}) can be used to determine the 
properties of the motion by using effective potentials.
The effective potential method is an advantageous tool used to understand and characterize the motion of  particles. Most importantly, this method allows us to infer the properties of the motion without any integration avoiding the numerical difficulties.

\subsection{General features of radial motion}
The radial motion for timelike particles moving along geodesics in the equatorial plane $\theta = \pi/2$, on a  
non-rotating Hayward black hole was described by~\cite{Chiba:2017nml, Hu:2018fle, Hu:2018old}. The rotating case in the equatorial plane was also considered  
in~\cite{Amir:2016nti}. Here, we summarize some of the results derived by~\cite{Amir:2016nti} 
in the equatorial plane by 
means of the effective potential method.

Taking the square of Eq.~(\ref{eq:forder_r}) and setting $Q=0$, the right hand side becomes a polynomial of second order in $E$. Hence, after some algebra one can rewrite the equation of motion for $r$ as 
\begin{eqnarray}
\left( \frac{d r}{d\tau} \right)^2
&=& -g^{tt}g^{rr}\left(E-V_{+}\right) \left(E-V_{-}\right) \ ,
\end{eqnarray}
where the potential functions $V_{\pm}$ in terms of the metric coefficients are

\begin{equation}
 V_{\pm}= \frac{g^{t\phi}}{g^{tt}}L_z \pm \left( 
\left[\left(\frac{g^{t\phi}}{g^{tt}}\right)^2-\frac{g^{\phi\phi}}{g^{tt}}  
\right]L_z^2- 
\frac{1}{g^{tt}} \right)^{1/2}.
\end{equation}
Substituting the expressions~(\ref{eq:inmetric_rothayward}) one 
gets:
\begin{equation}
 V_{\pm}= 	\frac{2aM(r)L_z}{r^3 + a^2(2M(r) + r)} \pm \left(\frac{\Delta [(r^2 + a^2)^2 - a^2 \Delta + r^2 L_z^2]}{[r^3 + a^2 
(2M(r)+r)]^2} 
\right)^{1/2} \ .
\label{eq:radpotential_delta}
\end{equation}

This expression reduces to Kerr in the limit for $M=\,$cte.~\cite{misner1970, Chandrasekhar1983}.
We can discuss the qualitative features of the motion 
of massive particles by plotting $V{\pm}$.
Physically acceptable orbits correspond to particles with energy $E$, greater 
than $V_{+}$. 

\begin{figure}[h!]
\begin{center}
\includegraphics[width=0.5\textwidth]
{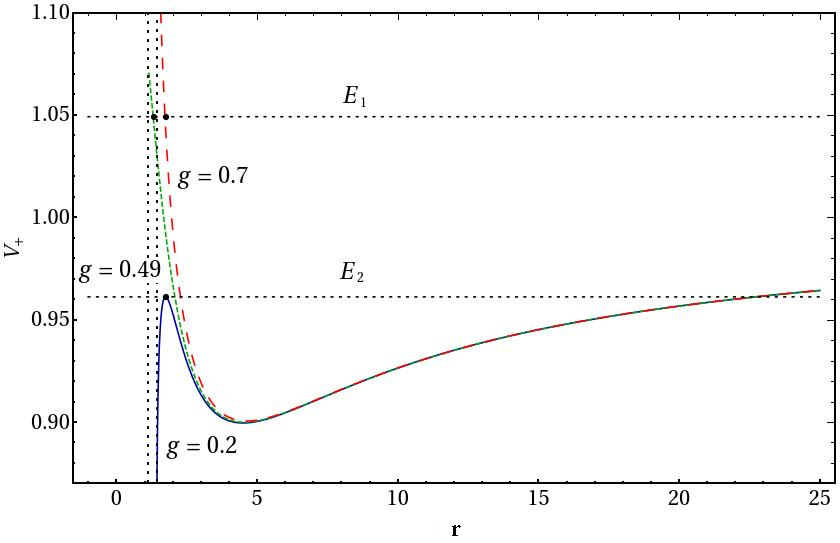}
\caption{Potential function $V_{+}$, for different values of $g$ with $a=0.9$ 
and $L_z=2.5$. For the potential with $g=0.7$ there is no horizon and as the 
values of $g$ decreases, the external horizon tends to the value given by the 
Kerr limit $r_{+_{Kerr}}=M+\sqrt{M^2-a^2}$.}
\label{fig:potential_gs}
\end{center}
\end{figure}

\begin{figure}[h!]
\begin{center}
\includegraphics[width=0.5\textwidth]
{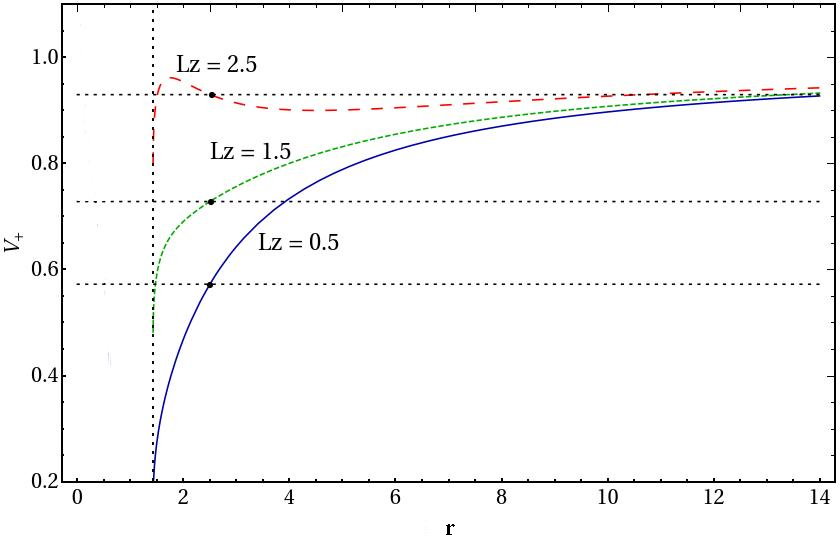}
\caption{Potential function $V_{+}$ for different values of  $L_z$. The black hole parameters are $g=0.2$ 
and
$a=0.9$. The vertical line refers to the external event horizon.
}
\label{fig:potential_ls}
\end{center}
\end{figure}

\begin{figure}[h!]
\begin{center}
\includegraphics[width=0.5\textwidth]{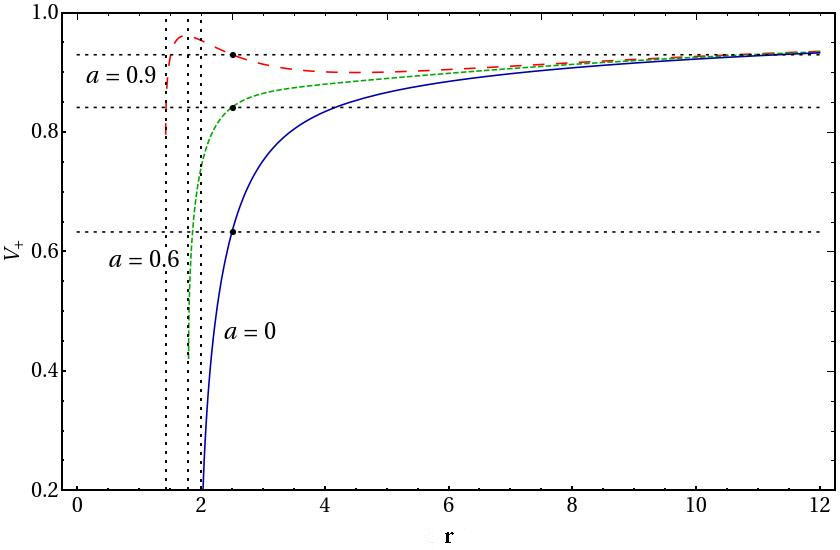}
\caption{Potential function $V_{+}$ for different values of $a$ and 
$g=0.2$. The angular momentum of the particle is $L_z=2.5$.}
\label{fig:potential_as}
\end{center}
\end{figure}

In Fig.~\ref{fig:potential_gs} are shown three different potential functions $V_{+}$, for spacetimes with no horizon, one horizon and 
two horizons obtained with three different values of the scale length 
$g=0.7,0.49,0.2$ respectively, with 
fixed angular momentum $L_z=3.0$ and $a=0.9$. 
The vertical dotted 
line at $r=1.13$ indicates the single event horizon for
$g=0.49$ while the line at
$r=1.42$ indicates the location of the external event 
horizon for $g=0.2$. For the last case there is a relative maximum in the potential indicating a 
(unstable) circular orbit.
Notice that for a fixed value of $a$, the external event horizon moves to smaller 
values of $r$ as the value of $g$ increases. For $a=0$ we recover the Hayward 
non-rotating space time~\cite{Hayward:2005gi}.
The potential functions for spacetimes with $g=0.49$ and $g=0.7$ are quite different from the case with $g=0.2$
In the former cases a particle with energy $E_1$ will reach a minimum radius and then escape towards infinity, while a particle with the same energy moving in a black hole with $g=0.2$ will fall.
A particle with energy $E_2$ cannot traverse the potential barrier for black holes with $g=0.49$ and $g=0.7$ but for a black hole with $g=0.2$ the particle will follow a circular orbit.
This different behaviour in the dynamics of the particles may be used to constraint the values of the parameter $g$.
In Fig.~\ref{fig:potential_ls} it is shown the 
potential for different values of $L_z$ with $g=0.2$ and 
$a=0.9$. For this case, the spacetime posses two horizons and
the values of the angular momentum are representative to show the 
behavior of the potential in the near horizon region.
The vertical dotted line indicates the position of the 
external horizon. As the angular momentum of the particle 
decreases, the potential develops a local maximum and a local minimum 
allowing unstable and stable circular orbits respectively. The particles with 
energy between the local maximum and minimum of $V_{+}$ 
have a bounded orbit whose radius lies between the turning points. In the limiting 
case of $L_z=0$ (radial motion)
the potential $V_{+}$ in Eq.~(\ref{eq:radpotential_delta}) becomes
%
monotone without a maximum or minimum. \\
From its derivative, the equation $V_{+}'=0$ has no real solutions for any value of the parameters, thus when $L_z = 0$ the particles will always fall into the black hole.  
In Fig.~\ref{fig:potential_as} it is shown the behavior of the potential for 
different values of the rotating parameter $a$. A general feature is that for 
large $r$ the curves are asymptotic to zero and, like in the Kerr black hole, 
the effect 
of rotation is only noticeable for small $r$. Furthermore, the effect of 
$g$ is also relevant for small values of $r$.

\begin{figure}[ht]
\centering	
\includegraphics[width=0.5\textwidth]{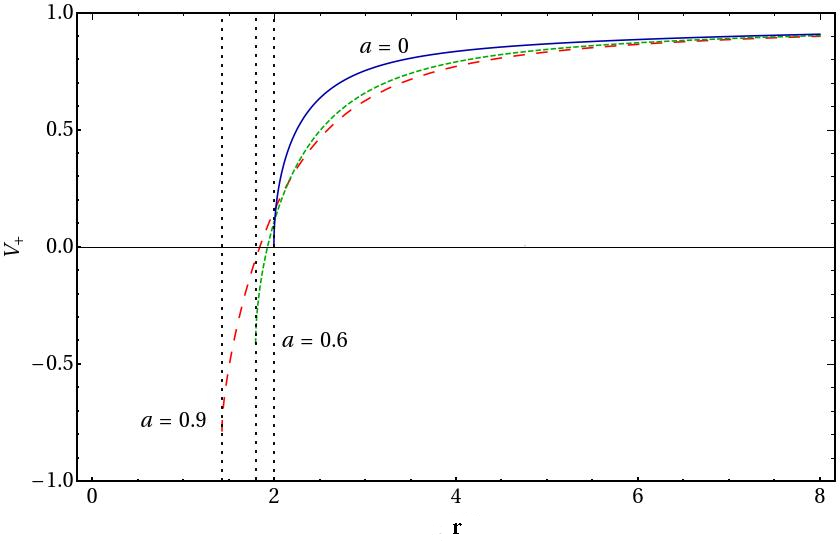}
\caption{Potential function, $V_{+}$ for $a=$  with 
$g=0.2$ and $Lz = -2.5$}
\label{fig:potential_a_g_02_lz_25_negative}
\end{figure}

A salient feature of the potential function $V_{+}$, is that for $aL_z < 
0$, it takes negative values as we 
approach the horizon. In the region where the potential is negative, the 
energy of the particles $E$ may be also negative and 
extraction of energy from the black hole is possible via the Penrose 
process.
In Fig.~\ref{fig:potential_a_g_02_lz_25_negative} we plot $V_{+}$  
for a negative value of the angular momentum $L_z=-2.5$ and for the rotation 
parameter $a=0.6$, $a=0.9$ and the limit $a=0$. 
The vertical dashed lines are the positions of the event 
horizons.
In the non rotating limit, the potential function is always positive, whereas 
for $a=0.6$ and $a=0.9$ there is a region close the horizon where motion with 
negative energy is allowed.
\subsection{General features of polar motion}
As compared to the orbits in the equatorial plane, polar motion has 
a richer variety of orbits.
A qualitative description of the motion can be performed analyzing the equation 
(\ref{eq:forder_th}). Let us introduce a new 
independent variable $u:=\cos\theta$. 
The square of the right hand side of 
equation~(\ref{eq:forder_th}) in terms of $u$ becomes a quartic polynomial
\begin{equation}\label{eq:dudt}
 \Sigma^2 \left(\frac{du}{d\tau} \right)^2 = f(u) = Q+ Au^2+B u^4
\end{equation}
where
\begin{equation}
 A:= -(Q+L_z^2-a^2(E^2-1)) \ , \qquad B := -a^2(E^2-1) \ .
\end{equation}
From~(\ref{eq:dudt}) we conclude that the motion is only possible for 
$f(u)\ge0$.
Furthermore, evaluating $f(u)$ and its derivatives $f'(u)$ and $f''(u)$, at the 
extreme values
\begin{equation}
 f(0) = Q,\qquad f(1) = -L_z^2 , \qquad f'(0)=0, \qquad f'(1)= 2(2B+A) , \qquad 
f''(0)=2A \ ,
\end{equation}
one infers that the particle can reach the axis 
$u^2=1$ ($\theta=0$ or $\theta=\pi$) if and only if $L_z=0$.

\begin{figure}[ht!]
\begin{center}
\includegraphics[width=0.5\textwidth]{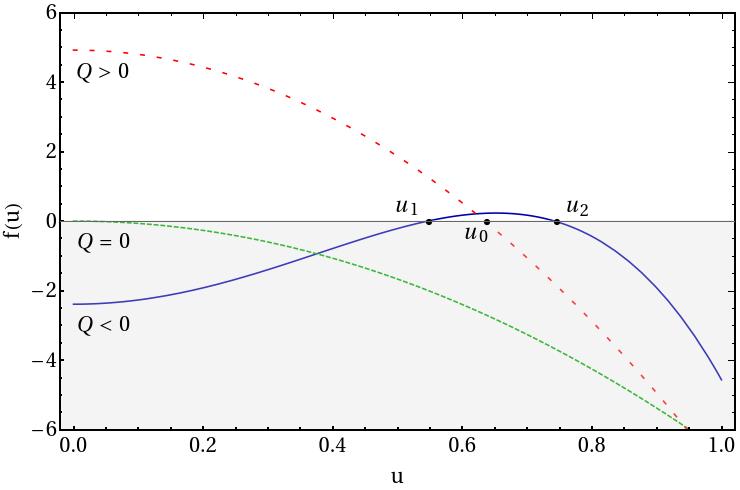}
\caption{Motion of particles is permitted in the region where $f(u)\ge0$. For $Q>0$  the values that $u$  can take are between $u=0$ and $u_0$, the movement is bounded and the particle can cross the equatorial plane. For $Q=0$ the particle remains in the equatorial plane $u=0$. For $Q<0$ the trajectory is bound between $u_1$ and $u_2$ but the particle cannot cross the equatorial plane.}
\label{fig:f_u}
\end{center}
\end{figure}

Following the analysis performed by Carter for the Kerr black 
hole~\cite{Carter:1968rr} we found the types of 
$\theta$ motion can be classified according to the sign of $Q$.
\begin{itemize}

\item If $Q<0$ then $f(0)<0$ and $f(1)\leq 0$, then the case of interest occurs 
when 
$f'(1)=(Q+L_z^2-a^2(E^2-1))^2+4a^2(E^2-1)Q\leq 0$.
In general the motion is oscillatory between $0<u_1\leq u \leq u_2$ where 
$u_1$ and $u_2$ are the two positive zeroes of $f(u)$. In this case however, 
the 
particle never crosses the equatorial plane. 

\item If $Q=0$ there is a trivial solution with $(E^2-1)=0=L_z$. Then $ 
\frac{du}{d\tau} =0$, 
and $\theta$ may take any constant value. There is a solution in which $\theta$ 
is constant at the equatorial plane $u=0$, ($\theta=\pi/2$) if 
$L_z<a^2(E^2-1)$. 
\item If $Q>0$ the particle moves in an oscillatory way crossing the equatorial 
plane with the angle $\theta$ lying in the range $\theta_0\leq \theta \leq 
\pi-\theta_0$ where $\cos\theta_0=u_0$
and $u_0$ is the unique zero of $f(u)$ in the range $0<u\leq 1$.
Aditionally, if $L_z=0$ and 
$Q-a^2(E^2-1)>0$, the particle will remain on the axis $\theta=0$. 
\end{itemize}
In Fig.~\ref{fig:f_u} we plot 
$f(u)$ with $Q=-2.0$, $Q=0$ and $Q=2.0$.
In this plot, we can identify the possible trajectories 
in the parameter space according to the behavior of $f(u)$. The motion is only possible in the non shaded region.

\section{Trajectories of particles in the configuration space}
\label{sec:trajectories}

In this section, we consider the trajectories of timelike particles in a Hayward rotating black 
hole in more detail.
The dynamics of test particles is governed 
by the geodesic 
equations~ (\ref{eq:forder_r}), (\ref{eq:forder_th}), (\ref{eq:forder_ph}), (\ref{eq:forder_t}). However, as 
described in \cite{Hughes:2001gg} for the 
Kerr spacetime, the 
square terms in $\dot r$ and $\dot \theta$ in the equations of motion  
cause problems in a numerical implementation when determining the turning 
points because there is a change of sign in $\dot r$ and $\dot \theta$ in 
those points. 
To deal with this problem we employ the Hamilton formulation to find the 
trajectories in the configuration space.
The hamiltonian and the equations of motion are
\begin{eqnarray}
 H =\frac{1}{2}g^{\alpha\beta}p_{\alpha}p_{\beta}\ .
\end{eqnarray}
and
\begin{equation}
\dot p_{\alpha} =- \frac{\partial H}{\partial x^{\alpha}}\ , \qquad 
\dot x^{\alpha} = \frac{\partial H}{\partial p_{\alpha}}\ .
\end{equation}
As before, the energy and the projection along the $z$ axis of the 
angular momentum are constant of motion 
$ p_t =-E$ and $p_\phi=L_z $. The equation 
for $\dot p_r$ is
\begin{eqnarray}
\dot p_r &=& -\frac{\partial H}{\partial r} = -\frac{1}{2}\left( g^{tt}_{,r}p_t^2+ 
g^{rr}_{,r}p_r^2+g^{\theta\theta}_{,r}p_{\theta}^2 + 
g^{\phi\phi}_{,r}p_{\phi}^2+2g^{t\phi}_{,r}p_{\phi}p_{t}   \right) \\
&=&-\frac{1}{2}\left( g^{tt}_{,r}E^2+ g^{rr}_{,r}p_r^2+g^{\theta\theta}_{,r}p_{\theta}^2 + 
g^{\phi\phi}_{,r}L_{z}^2-2g^{t\phi}_{,r}EL_z   \right) \ .
\label{eq:dotpr}
\end{eqnarray}
Using the normalization of the four momentum $p^{\mu}p_{\mu}=-1$ (recalling we have set the mass of the particles $\mu=1$) we can solve for  
$p_{\theta}^2$ 
\begin{equation}
 p_{\theta}^2= -\frac{1}{g^{\theta\theta}}\left(1+g^{tt}E^2+g^{rr}p_r^2+g^{\phi\phi}L_z^2-2g^{t\phi}EL_z\right) \ .
\end{equation}
Substituting this expression in~(\ref{eq:dotpr}) we get
\begin{eqnarray}
\dot p_r &=& \frac{1}{2} \left[
\left(  \frac{g^{tt}}{g^{\theta\theta}} g^{\theta\theta}_{,r}-g^{tt}_{,r}  
\right)E^2+
\left(  \frac{g^{rr}}{g^{\theta\theta}} g^{\theta\theta}_{,r}-g^{rr}_{,r}  
\right)p_{r}^2+
\left(  \frac{g^{\phi\phi}}{g^{\theta\theta}} 
g^{\theta\theta}_{,r}-g^{\phi\phi}_{,r}  \right)L_{z}^2-
2\left(  \frac{g^{t\phi}}{g^{\theta\theta}} 
g^{\theta\theta}_{,r}-g^{t\phi}_{,r} 
 \right)EL_{z}
 +  \frac{g^{\theta\theta}_{,r}}{g^{\theta\theta}}
 \right] \ .
 \label{eq:pr_dot}
\end{eqnarray}
This equation involves only radial derivatives of the metric coefficients and 
constants of motion.
The equation for $\dot p_{\theta}$ becomes
\begin{equation}
\dot{p_{\theta}} = -\frac{\partial H}{\partial  \theta}=-\frac{1}{2} (g^{tt}_{,\theta} 
E^2 + 2 
g^{t\phi}_{,\theta}E L_z + g^{rr}_{,\theta} p_r^2 + g^{\theta 
\theta}_{,\theta} p_{\theta}^2 + g^{\phi \phi}_{,\theta} L_z^2) \ ,
\label{eq:dotpth}
\end{equation}
Finally, in order to close the system, the equations for the coordinates in 
terms of the momenta and constants of motion are:
%

\begin{eqnarray}
 \dot t &=&  g^{t\phi} L_z - g^{tt} E \ ,  \label{eq:dot_t} \\
 \dot r &=& g^{rr}p_r \ , \label{eq:dot_r} \\
 \dot \theta &=& g^{\theta \theta} p_{\theta} \ , \label{eq:dot_th} \\
 \dot \phi &=&  g^{\phi \phi} L_z-g^{t \phi} E \ . \label{eq:dot_ph}
\end{eqnarray}

Note that since these equations do not contain square roots, they constitute 
a smoothly differentiable system of equations even at turning points and they 
can be integrated directly.
We solve equations (\ref{eq:dotpr}), (\ref{eq:dotpth}), (\ref{eq:dot_t}), (\ref{eq:dot_r}), (\ref{eq:dot_th}) and (\ref{eq:dot_ph}) numerically 
directly in the proper time using 
the 
\emph{Mathematica} numerical integrator NDSolve with a variable step-size Runge- 
Kutta integrator 
which, at each level takes a sequence of steps in the independent variable and uses an adaptive procedure to determine the size of these steps. 
NDSolve reduces the size step to track the 
solution accurately~\cite{mathematicaint}.

\subsubsection{Motion in the equatorial plane}

\begin{figure}[ht]
\centering
\includegraphics[scale=0.3]{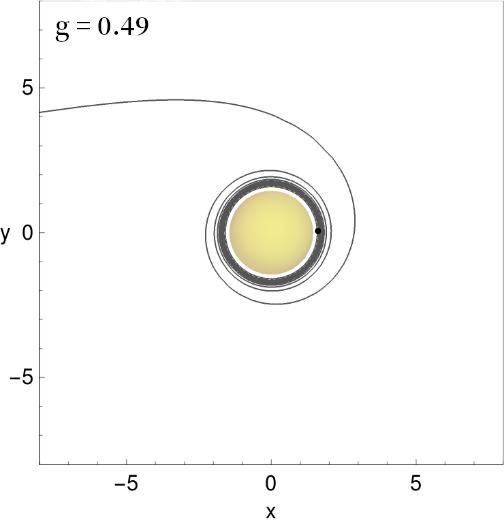}
\includegraphics[scale=0.3]{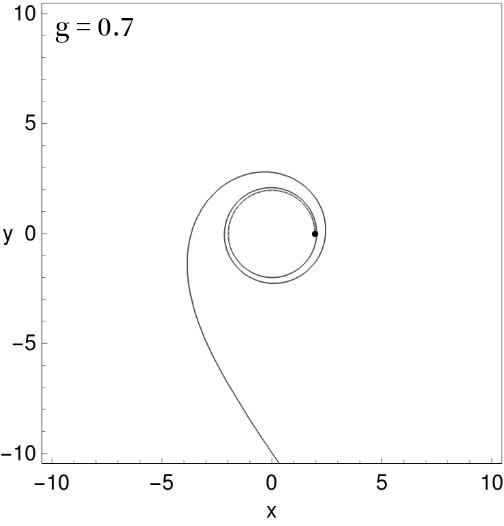}
\caption{Motion of a particle with energy $E_1=1.04$ for a spacetime with $g=0.49$ and $g=0.7$. In the former case the spacetime has one horizon and in the later there is no horizon.}
\label{fig:orb_a_09_g_s_lz_25e1}	
\end{figure}

\begin{figure}[ht]
\centering
\includegraphics[scale=0.3]{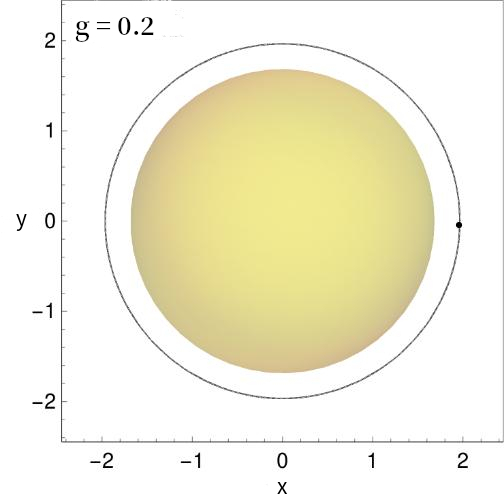}	
\includegraphics[scale=0.3]{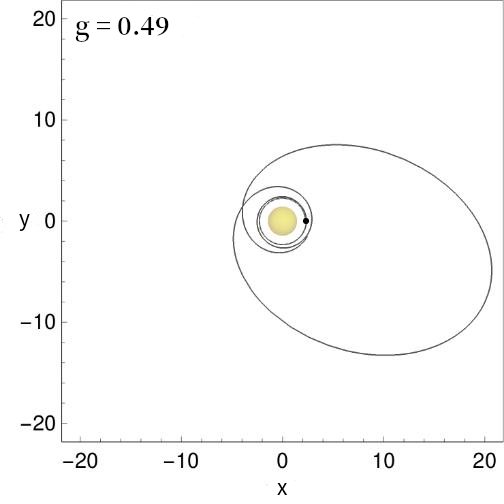}
\includegraphics[scale=0.3]{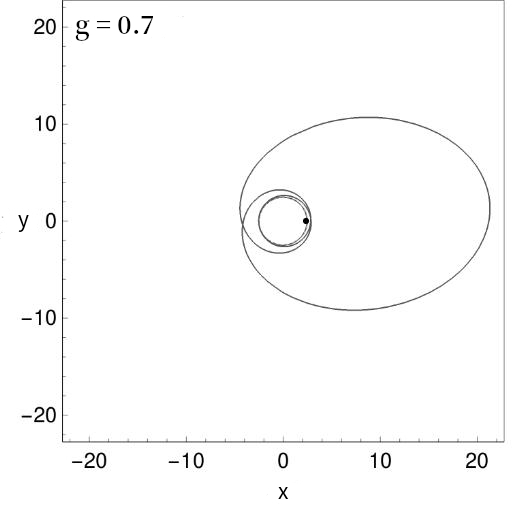}
\caption{Orbits for the radial potential with different values of $g$. The left 
panel corresponds to a spacetime with a single horizon, and the particle follows 
a (unstable) circular orbit. The central panel correspond to a spacetime with a 
single horizon and the trajectory of the particle is bound. The right panel 
shows the orbit of a particle moving on a spacetime with no horizons.}
\label{fig:orb_a_09_g_s_lz_25}	
\end{figure}

\begin{figure}
\centering	
\includegraphics[scale=0.3]{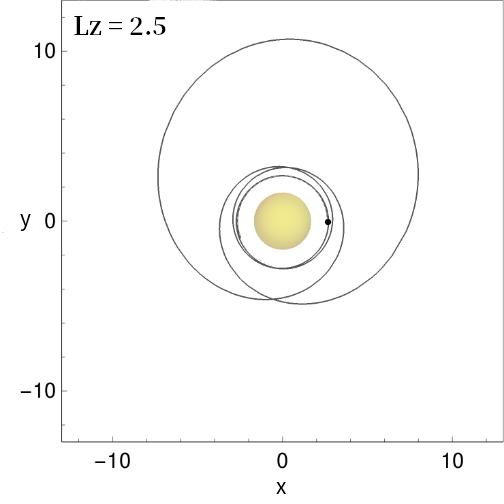}
\includegraphics[scale=0.3]{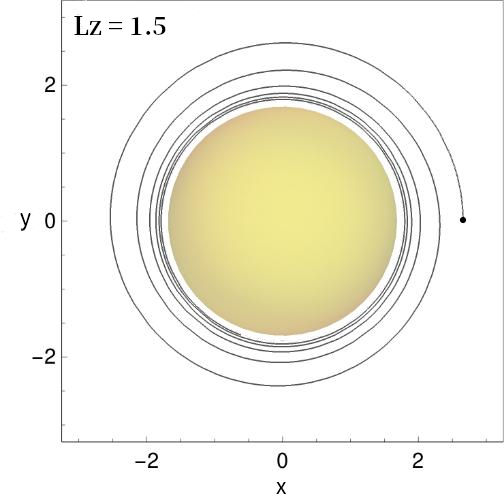}
\includegraphics[scale=0.3]{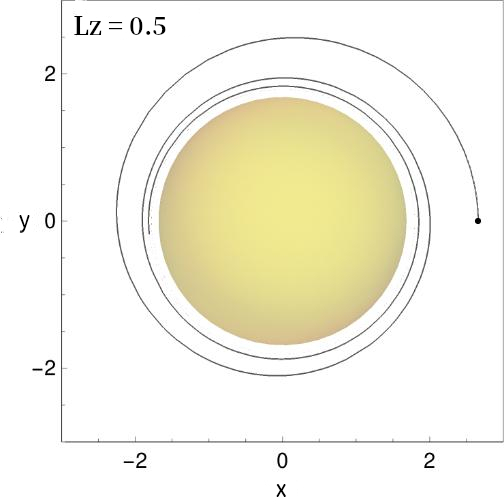}
\caption{Orbits for particles with different values 
of $Lz$. For the parameters $a=0.9$ and $g=0.2$ the metric corresponds to a 
black hole with two horizons.}
\label{fig:orb_a_09_g_02_lz_s}
\end{figure}

To start, let us focus on the radial motion of particles moving in the potential described in Fig.~\ref{fig:potential_gs}.
We solve the equations of motion numerically setting up the initial conditions in such a way that the particles have the energies $E_1$ and $E_2$ that correspond to the horizontal lines in Fig.~\ref{fig:potential_gs}.
In Fig.~\ref{fig:orb_a_09_g_s_lz_25e1}, we show the trajectory of a particle with energy $E_1=1.05$. As one can infer from the potential, 
the trajectory is unbounded for a spacetime with $g=0.49$ and a spacetime with $g=0.7$. Particles with energies $E>E_1$ in these spacetimes will reach a minimum radius and then escape to infinity. However, for a spacetime with $g=0.2$ the particle with the same energy will fall into the black hole.
In Fig.~\ref{fig:orb_a_09_g_s_lz_25}, we plot the trajectories of test particles 
for a rotating Hayward black hole with angular momentum $a=0.9$ and for
$g=0.2,0.49$ and $0.7$. The angular momentum of the particles is fixed at 
$L_z=2.5$ in all cases.
The particles have energy $E_2=0.961$, that correspond to 
the horizontal line $E_2$ in Fig.~\ref{fig:potential_gs}.
In Fig.~\ref{fig:orb_a_09_g_s_lz_25} (left panel) for $g=0.2$
the test particle moves on a unstable circular 
orbit with radius $r=1.98$, 
any small 
disturbance will make the test particle out of their original orbit. For the 
cases with $g=0.49$ and $g=0.7$ (middle and right panels) the particle with the 
same energy and angular momentum moves on non circular bound orbits.
For these orbits the precession direction is clockwise and the precession speed 
is faster than on the Kerr black hole with the same angular momentum. 
The orbits presented serve to exemplify the effect of the parameter $g$ on the trajectories of particles; for small values of $g$ a particle will fall into the horizon whereas a particle for larger values of $g$,
with the same initial conditions, will escape to infinity after encounter the potential barrier.
In Fig.~\ref{fig:orb_a_09_g_02_lz_s} we plot the trajectories of particles
on a 
rotating Hayward black hole 
with parameters
$a=0.9$ and $g=0.2$. The figures illustrate 
the trajectories for particles with angular momentum $L_z=2.5,\, 1,5,\, 0.5$. Like in 
the Kerr black hole, as the angular momentum of the particles decreases, the 
potential barrier vanishes yielding to trajectories that eventually fall into 
the black hole. 
The trajectories correspond to particles with constant energy 
represented by the horizontal lines in Fig.~\ref{fig:potential_ls}. 
The dots 
on  the intersection of the potential function and the energy in Fig.~\ref{fig:potential_ls} denote the 
initial position of the particles $r=2.5$ in all cases. The trajectory for the 
particle with 
$L_z=2.5$ is the only bound orbit with radius between $r=2.5 $ and $r=10.2$. For particles with $L_z=1.5$ and $L_z=0.5$ the trajectories are unbounded 
and the particles will fall into the black hole as can be infer from the potential function in Fig.~\ref{fig:potential_ls}.

\newpage

\begin{figure}[ht]
\centering	
\includegraphics[scale=0.3]{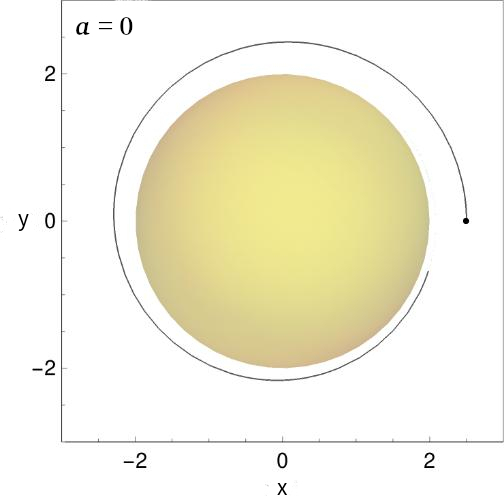}
\includegraphics[scale=0.3]{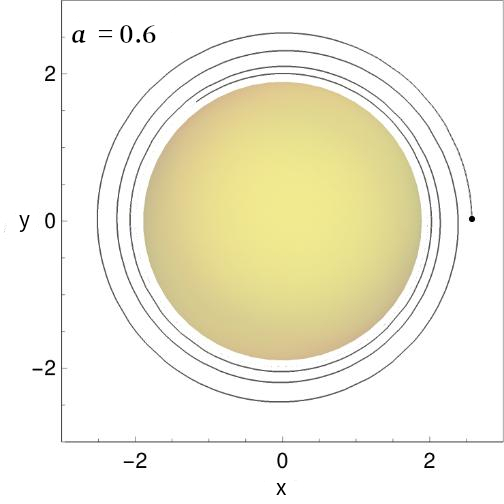}
\includegraphics[scale=0.3]{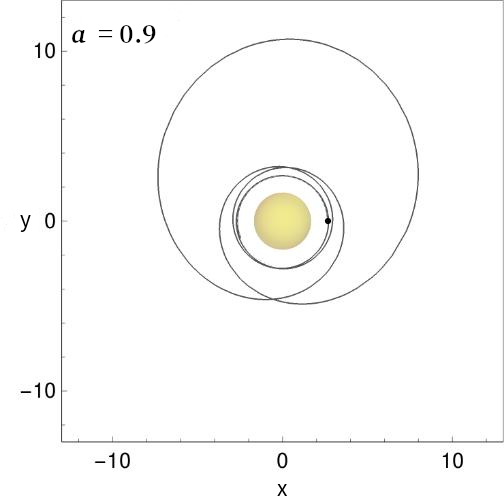}
\caption{Trajectories of particles orbiting a black hole  with different values of $a$. In each case the black hole
have two event horizons. For $a=0$ and $a=0.6$ the particle falls into the black hole
and for $a=0.9$ the trajectory is bounded between two radii.}
\label{fig:orb_a_s_g_02_lz_25}
\end{figure}

In Fig.~\ref{fig:orb_a_s_g_02_lz_25} we present the orbits of particles with  
angular momentum $L_z=2.5$, initial position $r_0 = 2.5$ which corresponds to the dots in the figure. The
constant energy is indicated for each particle with an horizontal line in Fig.~\ref{fig:potential_as}. The parameters of the black holes are $g=0.2$ and 
$a=0.9, 0.6, 0$. As illustrated in Fig.~\ref{fig:potential_as}, for the 
value $L_z=2.5$ the potential barrier vanishes as $a$ tends to zero and the local 
maximum and minimum of the potential tend to merge and eventually fade out for 
$a=0$. In this process, bound orbits disappear leaving only trajectories that fall into the black hole. 
For $a=0.6$ and $a=0$ the particles fall into the black hole while the particle moving around the  black hole with $a=0.9$ remains orbiting between two radii $r=2.5$ and $r=10.2$.

\begin{figure}[ht]
\centering	
\includegraphics[scale=0.3]{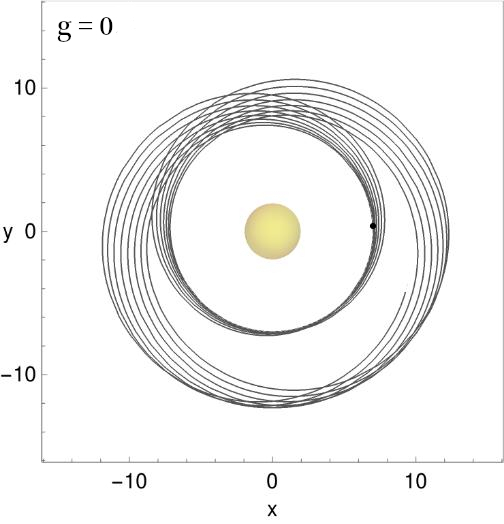}
\includegraphics[scale=0.3]{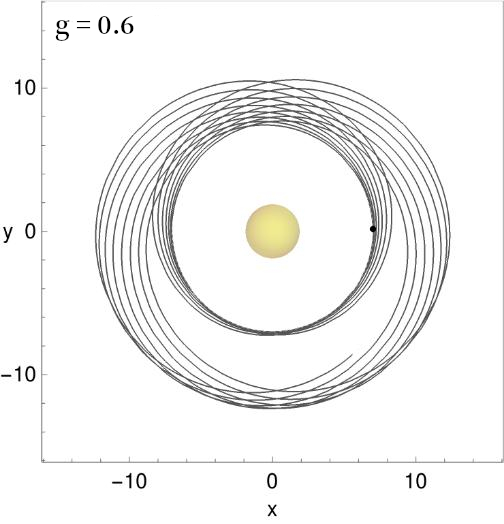}
\includegraphics[scale=0.3]{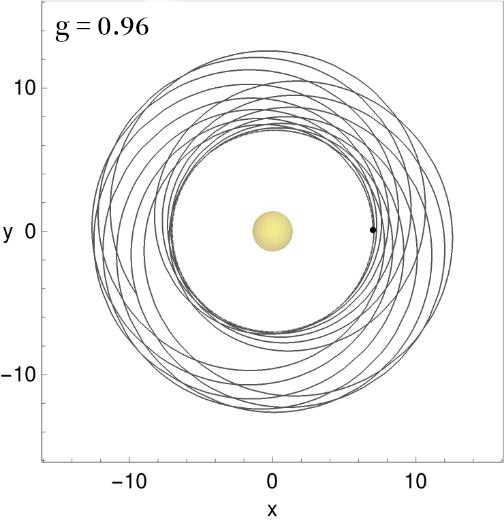}
\caption{Orbits of particles moving  in the equatorial plane around black holes with spin $a=0.4$.
The particles have angular momentum $L_z = 3.5$.
The motion in the equatorial plane corresponds to a Carter's constant $Q=0$.
}
\label{fig:orb_a_04_g_s_lz_35}
\end{figure}

In Fig.~\ref{fig:orb_a_04_g_s_lz_35} we show three plots for bound trajectories of particles moving around a black hole $g=0, 0.6$ and $g=0.96$ the last case corresponds to a limit value  in which the spacetime has one single horizon.
The spin 
of the black hole is $a=0.4$. The initial position is $r_0= 7.0$, the magnitude of the initial velocity is $v_0=0.45$ and the angular momentum of the particles is $L_z= 3.5$. The dots in Fig.~\ref{fig:orb_a_04_g_s_lz_35} corresponds to the initial position $r_0$. For these cases, the values of  $a$ and $L_z$ were chosen to show noticeable changes in the trajectories as $g$ varies. The time of evolution is $\tau = 2780$. 
As $g$ increases the difference in the trajectories becomes more noticeable.
For these cases, the change in the orbits with respect to Kerr black holes may become relevant for instance in the gravitational wave emission~\cite{Hughes:1999bq,Sasaki:2003xr}.
For instance when
in a binary system where binaries are assumed to consist of a Kerr massive black hole and a small compact star which is taken to be a point particle. 
The differences in the trajectories of massive particles in the equatorial plane, between the Kerr black hole and the rotating Hayward black hole, are the change in the angle of precesion. 
As $g$ increases, the precesion increases. 
This difference is more noticeable for small values of the spin parameter $a$, for nearly extreme black holes ($a \sim 1$) the differences become negligible.\\
Also we have found that the parameter $g$ may play an important role in the dynamics of massive particles, for a specific value of $a$ and the angular momentum, a particle would orbit the black hole or fall into it depending exclusively on the parameter $g$. \textit{E.g.} if a particle falls into a Kerr black hole, we can choose a value of $g$ in the rotating Hayward black hole such that the particle, with the same conditions, will not fall into the black hole.

\subsubsection{Motion out the equatorial plane}

Some studies of geodesics in the equatorial plane have been done for regular
black holes~\cite{Abbas:2014oua}. Nevertheless, there 
are not such studies for the motion \emph{out} the equatorial plane. 
It is important to make a fully study of the orbits out of the plane in order to make accurate comparisons
with the Kerr black hole and the data obtained from astronomical observations.
Here we calculate numerically the trajectories of particles moving on the rotating Hayward spacetime out the equatorial plane. As 
described above the orbits can be classified according the sign of the 
\emph{Carter constant} $Q$.

\begin{figure}[ht]
\centering	
\includegraphics[scale=0.3]{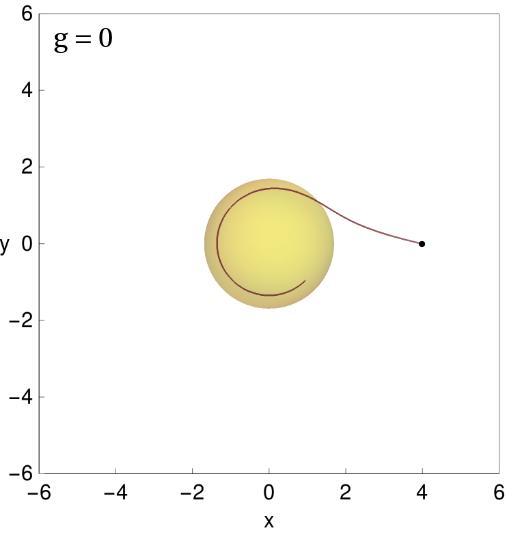}
\includegraphics[scale=0.3]{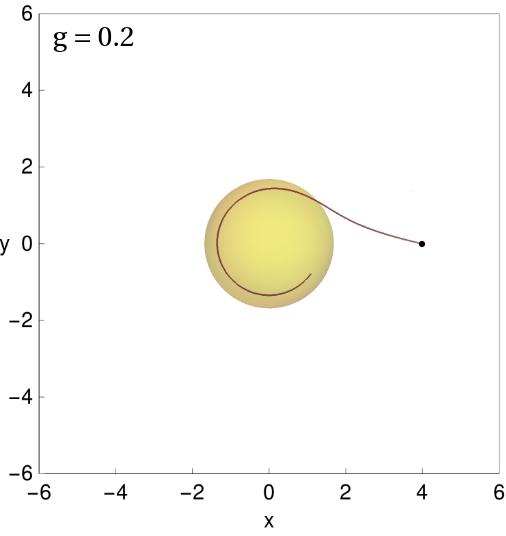}
\includegraphics[scale=0.3]{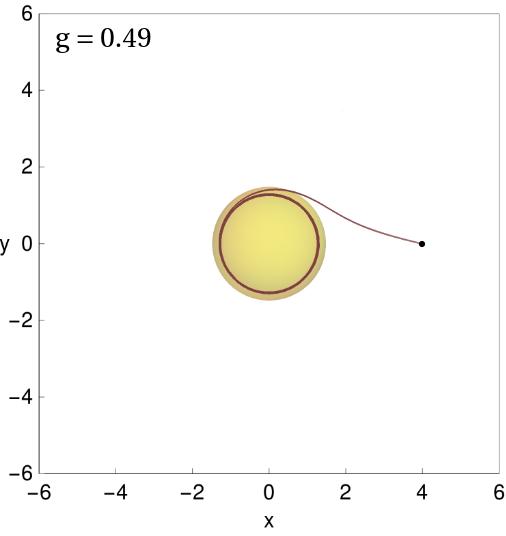}
\includegraphics[scale=0.3]{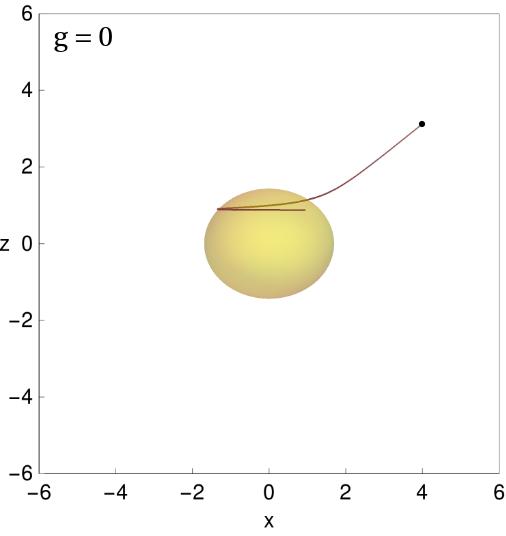}
\includegraphics[scale=0.3]{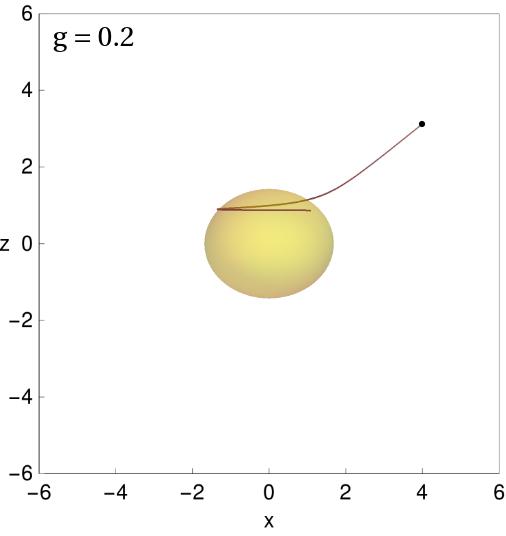}
\includegraphics[scale=0.3]{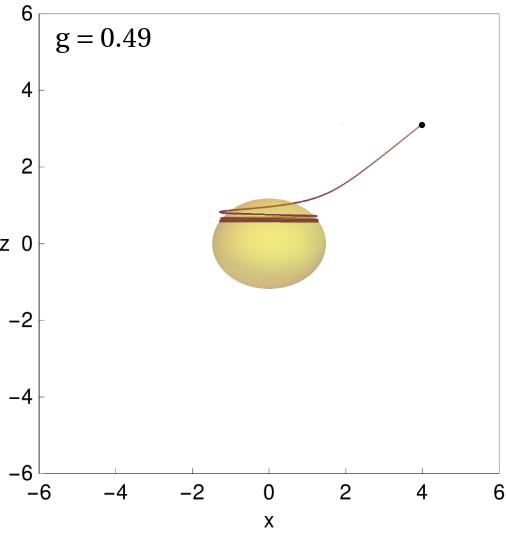}
\caption{Different views of the trajectories of particles moving in spacetimes with different $g$. The spin of the black hole is $a=0.9$ and the particles satisfy the condition $Q<0$. The figures in the right column correspond to a spacetime with a single horizon}
\label{fig:orbit_g_s_a_09}
\end{figure}

In Fig.~\ref{fig:orbit_g_s_a_09}
we plot the trajectories for a particle with a negative Carter's constant 
$Q=-0.2$. The two plots of the left panel corresponds to the 
Kerr spacetime with $g=0$, the plots in middle panel is a 
spacetime with two horizons $g=0.2$, and the right 
panel is a spacetime with a single horizon determined by $g=0.49$. 
In all cases the black hole has spin $a=0.9$.
For all particles, the initial velocity is $0.96$ and the initial position
$r_0 = 5.0$ and is represented with a dot in the plots. The time evolution in 
units of $M$ was $\tau=160$.
The first row corresponds to a projection of the orbit in the plane x-y. The 
black hole spin axis is z and is rotating counterclockwise. The second 
row corresponds to a projection in the z-x plane. 
The motion of the particle occurs between two angles $\theta_1$ and $\theta_2$, which
correspond to $u_1$ and $u_2$ of the function potential $f(u)$ in section~\ref{sec:geodesics}.
These angles change according to the value of $g$, as $g$ increases the difference 
between the two angles becomes slightly small.
For $g=0$ the value of $\theta$ lies between $\theta_1 = 0.316\pi$ and $\theta_2 = 0.233\pi$.
\newpage
\begin{figure}[ht]
\centering	
\includegraphics[scale=0.3]{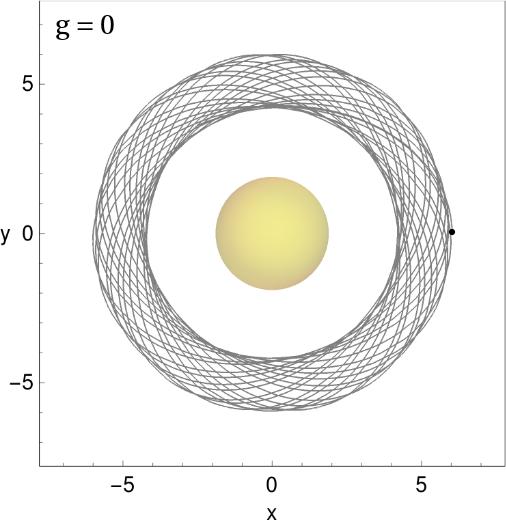}
\includegraphics[scale=0.3]{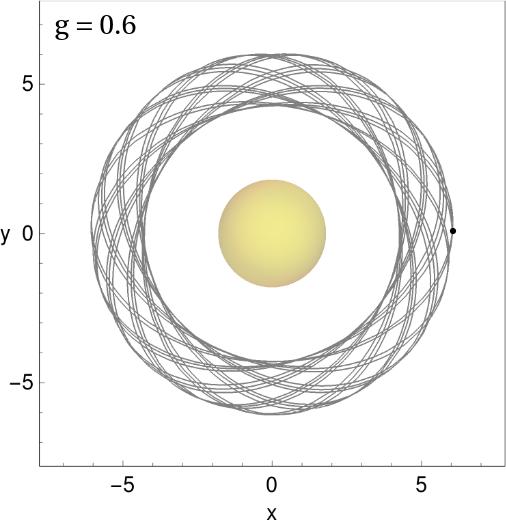}
\includegraphics[scale=0.3]{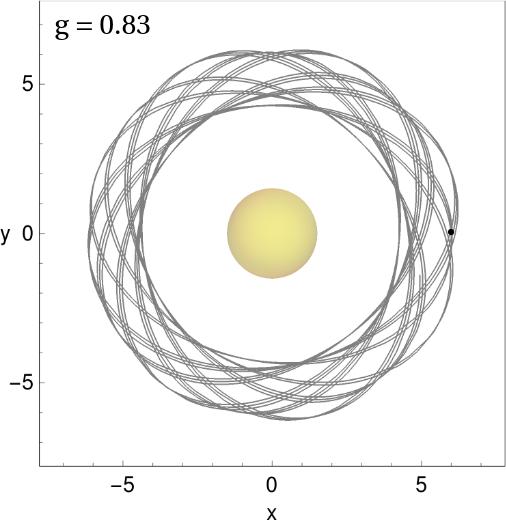}
\includegraphics[scale=0.3]{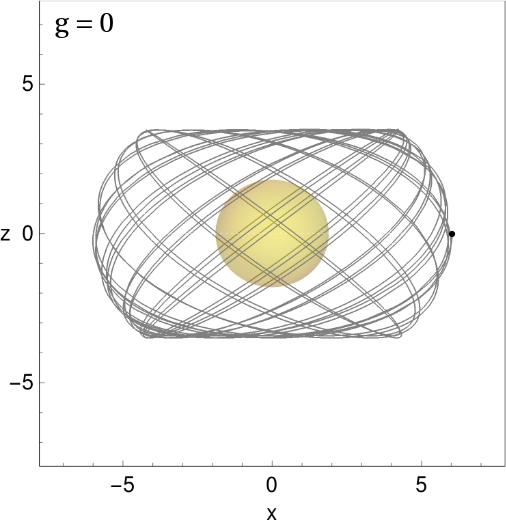}
\includegraphics[scale=0.3]{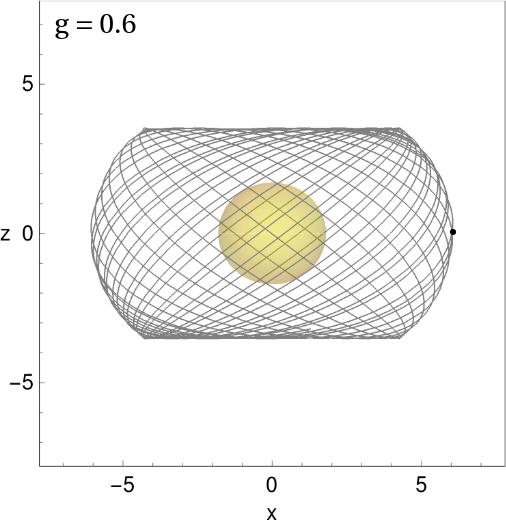}
\includegraphics[scale=0.3]{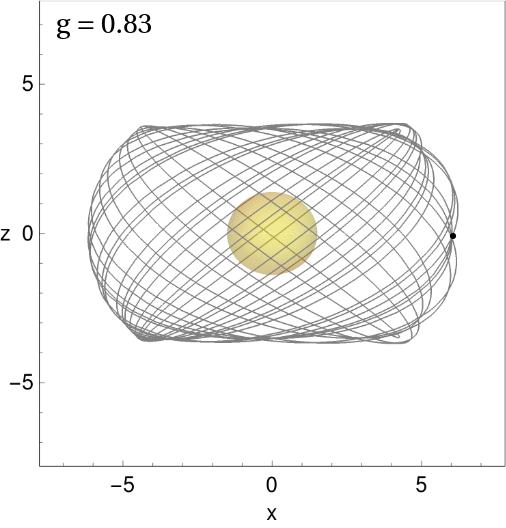}
\caption{Two views of the trajectories of particles moving around rotating black holes. The spin of the black holes is $a=0.6$. The spacetime with $g=0.836$ corresponds to a black hole with a single horizon.
The initial data of the particles were set such that the Carter's constant $Q>0$.}
\label{fig:orbit_g_s_a_06}
\end{figure}
In Fig.~\ref{fig:orbit_g_s_a_06}, the particles  initially satisfy the condition $Q>0$,
the initial velocity is $v_0=0.52$ and the initial position is $r_0 = 6$.
For the black hole with $g=0$, the particle is restricted to move between the angles 
$\theta_1 = 0.280\pi$ and $\theta_2 = 0.719\pi$.
For the cases illustrated, $g=0$ is the Kerr black hole and $g=0.6,\, g=0.8$ are rotating Hayward black holes with two and one horizon.
For rapidly rotating black holes, the effect of $g$ on the trajectories is 
almost negligible but for black holes with low spin, the trajectories change 
with respect to Kerr. As the value of $g$ increases the effect is more 
noticeable. 
However, there is a limit on $g$ in which the horizons of the 
black hole vanish yielding a spacetime with no horizons.

\section{Discussion and concluding remarks}
\label{sec:Conclusions}

In this work  we have described the properties of the horizons and ergospheres for a
rotating black hole geometry proposed by Bambi and Modesto in Ref.~\cite{Bambi:2013ufa}. The proposal was made as a generalization 
of the Hayward geometry \cite{Hayward:2005gi} and is characterized by a length scale $g$ and the spin $a$. 
We have studied the timelike geodesics around such rotating black holes. 
We discussed possible types of 
orbits in these spacetimes using effective potentials. We examined its properties paying particular attention to the changes due to the parameter $g$ the spin of the black hole $a$ and the angular momentum of the particles $L_z$.
We have described the properties of the geodesics using the Hamilton-Jacobi approach showing that for the rotating Hayward black hole proposed by Bambi and Modesto, there is a fourth constant of motion similar to the Carter constant in the Kerr black hole. We showed that the motion of particles out of the equatorial plane is well characterized in terms of this constant.
Furthermore, we found the trajectories of the particles integrating numerically the equations of motion considering particles lying \emph{in} and \emph{out} the equatorial plane.
We compare these trajectories with those of the Kerr black hole with the same mass and spin.
We found slight differences in the trajectories in both spacetimes. In particular, for slowly rotating black holes the differences become noticeable for large values of the parameter $g$. These differences may become relevant in studies of gravitational wave emission of compact objects orbiting massive black holes. 
It is the general belief that astrophysical black holes are of the Kerr type 
and the astrophysical importance of studies of regular black holes may have 
limited applications. However, the study of orbits around rotating black holes 
is important from conceptual and theoretical points of view since it is a 
medium to seek for differences among the black holes 
specially in the near horizon region where the parameters of the theory are important. 
Little is known about the orbits in
compact rotating objects other than Kerr. If such objects were to represent serious candidates for compact astrophysical objects, the presence of representative signatures would be highly valuable for their identification.


\acknowledgments
This work was supported in part by the CONACyT Network Project No. 294625 
``Agujeros Negros y Ondas Gravitatorias" and by 
UNAM-PAPIIT through 
grants IA101318 and IN104119. 
This work has further been supported by the European Union's Horizon 2020 research and innovation (RISE)
program H2020-MSCA-RISE-2017 Grant
No. FunFiCO-777740. 
BBO thanks CONACyT graduate program for support.



\end{document}